\documentclass[a4paper,11pt]{article}

\usepackage{jheppub} 

\usepackage[T1]{fontenc} 

\usepackage[bitstream-charter]{mathdesign}
\usepackage{upgreek}
\usepackage{xspace}
\usepackage[squaren]{SIunits}
\usepackage{comment}
\usepackage{caption}
\usepackage{xargs,xcolor,todonotes,soul}

\DeclareSymbolFont{usualmathcal}{OMS}{cmsy}{m}{n}
\DeclareSymbolFontAlphabet{\mathcal}{usualmathcal}

\newcommand{\vincia}{\textsc{Vincia}\xspace}
\newcommand{\madgraph}{\textsc{MadGraph5}\xspace}
\newcommand{\pythia}{\textsc{Pythia}\xspace}
\newcommand{\sherpa}{\textsc{Sherpa}\xspace}

\newcommand{\rivet}{\textsc{Rivet}\xspace}

\title{\boldmath Multi-Jet Production in Deep Inelastic Scattering with Pythia}

\author[a,b]{Ilkka Helenius,}
\author[a,b]{Joni O. Laulainen}
\author[c]{and Christian T. Preuss}

\affiliation[a]{University of Jyvaskyla, Department of Physics,  \\
P.O. Box 35, FI-40014 University of Jyvaskyla, Finland}
\affiliation[b]{Helsinki Institute of Physics, \\
P.O. Box 64, FI-00014 University of Helsinki, Finland}
\affiliation[c]{University of Wuppertal, Department of Physics, DE-42119 Wuppertal, Germany}

\emailAdd{ilkka.m.helenius@jyu.fi}
\emailAdd{joollaul@jyu.fi}
\emailAdd{preuss@uni-wuppertal.de}

\abstract{We introduce multi-jet merging for deep inelastic scattering in the \vincia parton shower in the Monte Carlo event generator \pythia~8. Merging combines event samples of different parton multiplicities with logarithmically enhanced parton-shower radiation. We consider up to five outgoing partons using two different merging algorithms. We vary the relevant scale choices and compare to experimental data by the H1 collaboration at the HERA collider. Results show that multi-jet merging improves the description to data, especially for low virtuality events, and that jet cross-sections converge when the partonic multiplicity exceeds the number of reconstructed jets.}

\arxivnumber{2410.20950}

\begin{document} 
\maketitle
\flushbottom

\section{Introduction}
\label{sec1:intro}
In the Standard Model of particle physics the interactions between quarks and gluons are described by the theory of Quantum Chromodynamics (QCD). The perturbative expansion of QCD has successfully predicted these interactions in various aspects, particularly in high-energy scatterings involving a small number of partons through fixed-order matrix-element (ME) calculations. These hard scattering-processes can be combined with parton showers (PS) which describe the perturbative emissions by evolving partons to low energy scales close to the confinement scale $\Lambda_\mathrm{QCD}$, where hadronization models are employed to generate realistic final configurations of particles. General purpose Monte Carlo (MC) event generators \cite{Buckley:2011ms} contain all these modelling and can thus provide complete event simulations of high-energy particle collisions.

Matrix-element calculations, however, get increasingly complicated and computationally demanding due to the intrinsic factorial scaling of the number of diagrams with the number of particles. Under the assumption that the radiation of additional particles appears at small relative angles (collinear) or at small relative energies (soft), radiative corrections can be modelled by parton showers, which evolve partons to lower energies through successive branchings. In traditional parton showers, these branchings are sampled according to the DGLAP-evolution equations \cite{Dokshitzer:1977sg,Gribov:1972ri,Lipatov:1974qm,Altarelli:1977zs}, which resum large logarithms arising from collinear emissions. Modern parton showers typically use some form of colour dipoles \cite{Sjostrand:2004ef,Nagy:2005aa,Schumann:2007mg,Winter:2007ye,Giele:2007di,Nagy:2007ty,Hoche:2015sya,Brooks_2020,Platzer:2009jq,Dasgupta:2020fwr,Forshaw:2020wrq,Preuss:2024vyu}, with angular-ordered showers providing a notable exception \cite{Gieseke:2003rz}. 

On the other hand, ME calculations accurately describe hard, well-separated jets. To consistently combine events from a fixed-order calculation with parton-shower emissions, it is necessary to avoid double- and undercounting of configurations modelled by these two approaches. When ME events with different number of hard partons are combined together and dressed with parton showers, the double counting can be avoided through multi-jet merging techniques which distribute the contributions of ME and PS descriptions to appropriate regions of phase space. Such a technique was first applied to electron-positron annihilation
\cite{Catani_2001} with a method called Catani-Krauss-Kuhn-Webber (CKKW) merging. Many improvements have been made since, notably the probabilistic consideration of all possible shower histories by Lönnblad in CKKW-L \cite{L_nnblad_2002} or preserving the unitarity of inclusive cross sections in UMEPS \cite{L_nnblad_2013,Platzer:2012bs,Bellm:2017ktr}. For a more detailed summary of some of the merging algorithms, see ref.~\cite{Lavesson_2008}.

In this study, we implement a multi-jet merging algorithm for deep inelastic scattering (DIS) events in \pythia \cite{bierlich2022comprehensive}, utilising the \vincia antenna shower \cite{Brooks_2020}. We employ the two merging algorithms mentioned above, CKKW-L and UMEPS, and consider two different options for the merging scale combined with appropriate factorization- and renormalization-scale choices. Multi-jet merging in DIS has previously been studied in the context of the \sherpa event generator in \cite{Carli_2010} which formed the basis of a NNLO matched calculation in \cite{H_che_2018} and of NLO multi-jet merging in \cite{knobbe2023nnlonll}. A number of dedicated \textsc{Powheg-Box} NLO+PS related works have recently been presented in \cite{Banfi_2024,Borsa:2024rmh,Buonocore:2024pdv,FerrarioRavasio:2024kem}. In earlier works \cite{ZEUS:1998uhq,ZEUS:1999nrh,Carli_2010} it has been shown that especially at low virtualities, a simplistic MC treatment of parton showers does not provide a good description of (multi-)jet final states. In light of this, the presented merging implementation will ensure accurate predictions of DIS jet measurements for the forthcoming Electron-Ion Collider (EIC) \cite{ABDULKHALEK2022122447} and can be used to revisit diffractive dijet production cross sections measured at HERA to further study the observed breakdown of factorization \cite{Helenius:2019gbd}.

In section~\ref{sec2:theory}, we outline the theoretical background, covering the DIS process and its relevant physical quantities, parton showers, and multi-jet merging in the context of Monte Carlo event generators. Section~\ref{sec3:setup} expands on the details of our computational setup, with a focus on the significance of scale choices, and also contains intermediate validation results and sanity checks. In section~\ref{sec4:results}, we compare our implementation to experimental data. Finally, the conclusions and overall outcomes of the project are discussed in section~\ref{sec5:conclusion}.

\section{Theoretical background}
\label{sec2:theory}
Generating realistic collision events numerically is an elaborate task, which can be divided into many independent computations. One of the fundamental divisions is justified by the factorization theorem \cite{Collins:1989gx}, which motivates to separate short and long-distance physics. It allows to express the long-distance particle-level cross section as a convolution of the short-distance parton-level cross section governing high-energy physics and universal parton distribution functions (PDFs) which determine low-energy physics. The simulation of scattering events proceeds by first computing matrix elements to generate a core $2 \rightarrow N$ process according to the partonic cross section, and then applying a parton shower to model further branchings from these particles.
The primary objective of multi-jet merging is to consistently combine multiple higher-multiplicity fixed-order matrix-element calculations with parton showers. Events generated according to fixed-order MEs effectively capture the characteristics of hard, well-separated jets, while parton showers offer a suitable description of soft and collinear partons, describing jet substructure.
In multi-jet merging the PS approximation is assigned to the phase space region where it is most suitable, particularly for near-unresolved emissions and have MEs describe resolved partons \cite{Catani_2001,L_nnblad_2002}.

\subsection{Parton showers}
\label{sec2.1:partonshowers}
Parton shower algorithms are a crucial part of Monte Carlo event generation. They make use of universal splitting functions to describe the evolution of a scattering event in some energy scale and are therefore intrinsically process-independent.
The evolution is governed by the parton-shower evolution variable, which, in broad terms, represents the hardness of the event. 
In doing so, logarithmically enhanced terms are resummed.
The formal accuracy of the PS resummation may be expressed in terms of a power series in logarithms, as leading-logarithmic (LL), next-to-leading-logarithmic (NLL) and so on. 
In the present study we can generally only claim formal LL correctness, although there are recent developments on NLL-consistent parton shower algorithms \cite{Bewick:2019rbu,Dasgupta:2020fwr,Forshaw:2020wrq,Herren:2022jej,Bewick:2021nhc,vanBeekveld:2022ukn,vanBeekveld:2022zhl,Assi:2023rbu,Hoche:2024dee,Preuss:2024vyu} and beyond \cite{FerrarioRavasio:2023kyg,vanBeekveld:2024wws}, also in the context of DIS processes \cite{vanBeekveld:2023chs}. 

The default parton shower algorithm in \pythia is the transverse-momentum-ordered ``simple shower'' algorithm \cite{Sjostrand:2004ef}. 
Given that the simple shower is known to not faithfully reproduce coherent branchings in initial-final dipoles \cite{Jager:2014vna,Rauch:2016upa,Jager:2020hkz,Covarelli:2021gyz,Hoche:2021mkv}, it is in particular not suitable for DIS topologies. 
An extension to introduce dipole-antenna-like initial-final radiation patterns into the evolution has been given in \cite{Cabouat:2017rzi}, which includes a recoil handling appropriate for coherent branchings in DIS.
In the simple shower, branching probabilities are obtained from DGLAP splitting kernels $\hat{P}_{a\rightarrow bc}(z)$. 
The probability for a parton $a$ to split into partons $b$ and $c$ can thus be formally written as
\begin{equation}
    \mathrm{d} \mathcal{P}_a = \frac{\mathrm{d}t}{t} \frac{\alpha_\mathrm{S}}{2\pi} \sum_{b,c =\{q, g\}} \hat{P}_{a\rightarrow bc}(z) \mathrm{d}z,
\end{equation}
where $t$ is the parton-shower evolution variable and $z$ the energy fraction of parton $b$. Parton showers employ the veto algorithm to produce emissions.  

A full-fledged antenna shower in \pythia~8.3 is given by \vincia \cite{Brooks_2020}. In the \vincia framework, the evolution of partons is described in terms of the antenna formalism, in which partons are coherently emitted from two (leading-colour) dipole ends, $IK\to ijk$,
\begin{equation}
    \mathrm{d}\mathcal{P}_{j/IK} = \frac{\alpha_\mathrm{S}}{2\pi} \sum\limits_{j = \{q , g\}}C_{j/IK}\, A_{j/IK}(t,\zeta)\, \mathrm{d}t\, \mathrm{d}\zeta\, \frac{\mathrm{d}\phi}{2\pi} \, ,
\end{equation}
where $C_{j/IK}$ is the (leading) colour factor associated to the branching and $\zeta$ denotes an auxiliary phase-space variable while $\phi$ is the azimuthal angle.
Branching probabilities are described in terms of antenna functions rooted in the soft eikonal,
\begin{equation}
    A_{j/IK}(p_i,p_j,p_k) = \frac{2s_{ik}}{s_{ij}s_{jk}} + \mathrm{collinear~terms} \, .
\end{equation}
This ensures that, at least at leading colour, \vincia faithfully accounts for the coherent suppression of soft-gluon radiation in initial-final (IF) dipoles, as encountered in DIS and vector-boson-fusion (VBF) processes \cite{Cabouat:2017rzi,Hoche:2021mkv}. This has also been addressed in other studies regarding to VBF \cite{Jager:2020hkz,Ballestrero:2018anz}.
Parton branchings are simulated in \vincia as an interleaved sequence ordered in transverse momentum defined as
\begin{equation}
    t \equiv p_\perp^2 = \frac{\bar{q}_{ij}\bar{q}_{jk}}{s_\mathrm{max}} \, .
\end{equation}
Here, the invariant $\bar{q}_{ij}$ is defined as
\begin{equation}
    \bar{q}_{ij} = \begin{cases}
    (p_i+p_j)^2-m_I^2 & i \text{ final} \\
    -(p_i-p_j)^2+m_I^2 & i \text{ initial}
    \end{cases} \, ,
\end{equation}
and $s_\mathrm{max}$ denotes the largest invariant in the dipole
\begin{equation}
    s_\mathrm{max} = \begin{cases}
        s_{IK} & \text{FF} \\
        s_{ij} + s_{jk} & \text{IF} \\
        s_{ik} & \text{II}
    \end{cases} \, .
\end{equation}
For final-final (FF) configurations, the post-branching kinematics are constructed using a local recoil scheme, in which the transverse recoil is shared between two antenna parents $I$ and $K$. In configurations where at least one parent is in the initial state, such as initial-final or initial-initial (II) dipoles, initial-state particles are constrained to the beam axis and only acquire a longitudinal recoil. The transverse recoil is then absorbed by the final-state parent or the entire final state in case the other parent is a final-state parton or an initial-state parton, respectively. 

In the current version, \vincia resorts to the so-called sector-shower formalism, in which branchings are restricted to non-overlapping sectors in phase space \cite{Lopez-Villarejo:2011pwr,Brooks:2020upa}. This has a number of consequences. First and foremost, each antenna function is required to reproduce the full tree-level singularity structure associated to its phase-space sector, meaning that both the full soft and the full collinear limit is contained in the branching kernel. Secondly, this implies that, at least for gluon emissions at leading colour, the shower sequence is uniquely invertible while fully respecting the quantum nature of the evolution process. In the presence of gluon splittings, all possible quark-antiquark clusterings have to be taken into account. This has the effect that sector showers are ``maximally bijective'', that is, they produce the smallest possible numbers of branching sequences to any given phase-space point.
In particular, this facilitates the most efficient multi-jet merging algorithm \cite{Brooks:2020mab}.

\subsection{Multi-jet merging}
\label{sec2.2:merging}
To accomplish multi-jet merging without double-counting, a distinction must be made between hard and soft partons in order to divide the phase space into ME and PS regions. This is done by introducing a merging scale denoted by $t_\mathrm{MS}$, a kinematic quantity often defined in terms of the shower evolution variable, but can in principle be any partonic scale. 
We define
\begin{align*}
    &\text{hard partons: } t > t_\mathrm{MS}
    &\text{soft partons: } t \leq t_\mathrm{MS}.
\end{align*}
Here, $t$ represents the jet resolution variable, which is closely related to the shower evolution variable since shower splitting scales correspond to jet clustering scales in the merging algorithm. We also define the term $n$-parton sample to refer to the parton-level event sample, and $n$-jet sample to refer to the weighted event sample after merging takes place. 
Following \cite{L_nnblad_2013}, let us use the following notation for the hard process $+n$-jet cross section,
\begin{align}
\label{eq2.7:B_n}
     \mathrm{B}_n = \frac{\mathrm{d}\sigma_n^\mathrm{ME}}{\mathrm{d}\Phi_n} &= \underbrace{f_n^+(x_n^+, \mu_\mathrm{F}) \, f_n^-(x_n^-, \mu_\mathrm{F})}_{f_n(x_n, \mu_\mathrm{F})} \, |\mathcal{M}_n(t_n, \mu_\mathrm{F})|^2,
\end{align}
where $f_n^\pm$ are the beam PDFs, $\mathcal{M}_n$ the matrix element of the process with $n$ additional partons and $\Phi_n$ the corresponding phase space element. Let ($\mathcal{S}_0, \mathcal{S}_1, \dots, \mathcal{S}_n)$ be the states with $0 \dots n$ additional partons, ($t_0, t_1, \dots, t_n)$ the corresponding scales at which state $\mathcal{S}_i$ branches to $\mathcal{S}_{i+1}$, $0\leq i < n$, $\mathcal{P}_{\mathcal{S}_n}(t)$ the branching probability of a state and 
\begin{equation}
    \label{eq2.8:Sudakov}
    \Pi_{\mathcal{S}_n}(t_1, t_2) = \exp\Bigg( - \int_{t_2}^{t_1} \mathrm{d}t \, \alpha_\mathrm{S}(t) \, \mathcal{P}_{\mathcal{S}_n}(t) \Bigg)
\end{equation} 
the no-emission probability of state $\mathcal{S}_n$ between scales $t_1$ and $t_2$. The following sections are written with simplified notation, considering only final-state radiation (FSR) of parton showers while omitting initial-state radiation (ISR), and taking implicit integrals in $z$. If we attach the parton shower to the core process $\mathrm{B}_0$, the exclusive parton-shower cross-sections become
\begin{align}
    \frac{\mathrm{d}\sigma_0^\mathrm{PS}}{\mathrm{d}\Phi_0} &= \mathrm{B}_0 \Pi_{\mathcal{S}_0}(t_0,t_c) \\    \frac{\mathrm{d}\sigma_1^\mathrm{PS}}{\mathrm{d}\Phi_0} &= \int_{t_c}^{t_0} \mathrm{d}t \mathrm{B}_0 \Pi_{\mathcal{S}_0}(t_0,t) \mathcal{P}_{\mathcal{S}_0}(t) \Pi_{\mathcal{S}_1}(t, t_c) \\
    \label{eq2.11:sigmaPS}
    \frac{\mathrm{d}\sigma_2^\mathrm{PS}}{\mathrm{d}\Phi_0} &= \int_{t_c}^{t_0} \mathrm{d}t \mathrm{B}_0 \Pi_{\mathcal{S}_0}(t_0,t) \mathcal{P}_{\mathcal{S}_0}(t) \int_{t_c}^{t} \mathrm{d}t_1  \Pi_{\mathcal{S}_1}(t,t_1) \mathcal{P}_{\mathcal{S}_1}(t_1) \Pi_{\mathcal{S}_2}(t_1,t_c),
\end{align}
where $t_c \sim \Lambda_\mathrm{QCD}^2$ is the parton-shower cutoff. 

\begin{figure}
 \centering
    \includegraphics[scale=1.15]{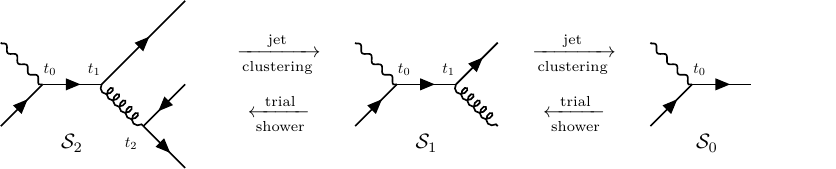}
 \caption{Illustration of how a merging algorithm reconstructs shower histories and produces shower states. }
 \label{fig1:mergingIllustration}
\end{figure}

With this notation at hand, let us write the effect of a general merging algorithm to the exclusive $+2$-jet cross section. Assume that a matrix element generator generates events according to the cross sections $\mathrm{B}_n$ with all partons above the merging scale, $t > t_\mathrm{MS}$, and we attach the parton shower to these samples with emissions below $t_\mathrm{MS}$. The integrated exclusive $+2$-jet cross section, that is, the cross section for exactly two additional jets, can then be written as 
\begin{equation}
\begin{split}
\sigma_2^\mathrm{ME+PS} = \int \mathrm{d}\Phi_0 \Bigg[ &\int_{t_c}^{t_0}\mathrm{d}t \mathrm{B}_0 \Pi_{\mathcal{S}_0}(t_0,t) \mathcal{P}_{\mathcal{S}_0}(t) \int_{t_c}^{t} \mathrm{d}t_1 \Pi_{\mathcal{S}_1}(t,t_1) \mathcal{P}_{\mathcal{S}_1}(t_1) \Pi_{\mathcal{S}_2}(t_1,t_c) \\
+ &\int_{t_c}^{t_1} \mathrm{d}t \mathrm{B}_1 \Pi_{\mathcal{S}_1}(t_1,t) \mathcal{P}_{\mathcal{S}_1}(t) \Pi_{\mathcal{S}_2}(t,t_c) \\
+ &\int_{t_c}^{t_2} \mathrm{d}t \mathrm{B}_2 \Pi_{\mathcal{S}_2}(t,t_c) \Bigg].
\end{split}
\label{eq2.12:NaiveMerge2jetSigma}
\end{equation}
However, even with the merging scale cut, there is overlap between the different event classes in this description.
For instance, both the first and the second line of eq. \eqref{eq2.12:NaiveMerge2jetSigma} contribute branchings with $t < t_\mathrm{MS}$.
The overlap can be avoided by making these inclusive ME samples exclusive, first by determining a possible parton-shower history of the given ME state; then by reweighting with factors that would have been present if the parton shower actually generated the state. 
This can be achieved by multiplying the cross sections $\mathrm{B}_n$ with a merging weight $w_n$. In practice, the merging weight is calculated by reconstructing all possible sets of shower states $\mathcal{S}_n$ and corresponding scales $t_n$ and generating a trial shower between all nodes in the most probable shower history, as is illustrated in figure \ref{fig1:mergingIllustration}. In trial showers, branchings are restricted to the region $t < t_n$.
This ensures that partons in higher-multiplicity events acquire the Sudakov suppression that would have been generated by the no-branching probability of the parton shower.
After the trial showers terminate, the main shower may produce further emissions with $t < t_\mathrm{MS}$ and the event generation proceeds as usual. 

In the context of the implementation in \pythia, we consider the CKKW-L scheme, suitably modified for DIS processes.
The way it introduces higher-order corrections is of additive nature, which results in a non-unitary merging scheme \cite{Lavesson_2008}. The effect of the algorithm on a given event is encapsulated in the CKKW-L weight 
\begin{equation}
    w_n = w^{\mathrm{PDF}_0} \, w^{\mathrm{PDF}_i} \, w^{\alpha_\mathrm{S}} \, w^\mathrm{Sudakov} ,
\end{equation}
where
\begin{align}
    \nonumber
    w^{\mathrm{PDF}_0} &= \frac{f_0(x_0, t_0)}{f_n(x_n, \mu_\mathrm{F})} & w^{\mathrm{PDF}_i} &= \prod_{i=1}^n \frac{f_i(x_i, t_i)}{f_{i-1}(x_{i-1}, t_i)} \\ 
    w^{\alpha_\mathrm{S}} &= \prod_{i=1}^n \frac{\alpha_\mathrm{S}(t_i)}{\alpha_\mathrm{S}(\mu_\mathrm{R})} & w^\mathrm{Sudakov} &= \Pi_{\mathcal{S}_{n}}(t_{n}, t_\mathrm{MS}) \, \prod_{i=1}^n \Pi_{\mathcal{S}_{i-1}}(t_{i-1}, t_i). 
\end{align}
Factors containing the strong coupling constant ensure that the PS and ME contributions account for the running of $\alpha_\mathrm{S}$ similarly. The first PDF ratio takes care of normalization to the same Born level cross section, by substituting $\mathrm{B}_n^\mathrm{ME} = f_n(x_n,\mu_\mathrm{F})|\mathcal{M}_n|^2 \rightarrow \frac{f_0(x_0,t_0)}{f_n(x_n,\mu_\mathrm{F})} \mathrm{B}_n^\mathrm{ME} = f_0(x_0,t_0)|\mathcal{M}_n|^2$. The other PDF ratios of $w^{\mathrm{PDF}_i}$ handle normalization required by backwards parton evolution, and no-branching probabilities provide the Sudakov suppression according to the parton shower at hand. 

The CKKW-L procedure can be written algorithmically in the following way for $+N$-jet merging: 
\begin{enumerate}
    \item Start from an event from the fixed-order matrix element sample. Enforce the merging scale cut, so that each additional parton obeys $t > t_\mathrm{MS}$. 
    \item Construct shower histories by inverting the shower, obtain states $\{\mathcal{S}_0, \mathcal{S}_1 \dots \mathcal{S}_n \}$ and corresponding scales $\{t_0, t_1 \dots t_n \}$. Choose the history probabilistically. 
    \item Perform trial emissions from states $0 \leq i < n$. Reject the event if the generated emission is above $t_{i+1}$.
    This step has the same effect as multiplying with the corresponding no-branching probabilities. Reweight the event with PDF ratios and $\alpha_\mathrm{S}$ ratios. 
    \item Start the parton-shower cascade from $t_n$.
    \begin{itemize}
        \item if $n<N$: reject the event if the first emission is above $t_\mathrm{MS}$, otherwise accept it;
        \item if $n=N$: accept the event.
    \end{itemize}
    If the event has been accepted, run the shower down to $t_\mathrm{c}$.
\end{enumerate}
When using \vincia's sector merging, the number of histories is one when only leading-colour gluon emissions are present and scales at most factorially with the number of same-flavour quark pairs when gluon splittings are present.

If we constrain the merging to the case with two additional jets, the algorithm reweights different contributions and modifies the cross sections as
\begin{align}
    \frac{\mathrm{d}\sigma_0^\mathrm{ex}}{\mathrm{d}\Phi_0} &= \mathrm{B}_0 \Pi_{\mathcal{S}_0}(t_0, t_\mathrm{MS}) = w_0 \mathrm{B}_0 \\
    \frac{\mathrm{d}\sigma_1^\mathrm{ex}}{\mathrm{d}\Phi_0} &=   \frac{f_1(x_1,t_1)}{f_0(x_0,t_1)} \frac{\alpha_\mathrm{S}(t_1)}{\alpha_\mathrm{S}(\mu_\mathrm{R})} \Pi_{\mathcal{S}_0}(t_0, t_1) \Pi_{\mathcal{S}_1}(t_1, t_\mathrm{MS}) \, \mathrm{B}_1 = w_1 \mathrm{B}_1 \\
    \frac{\mathrm{d}\sigma_2^\mathrm{in}}{\mathrm{d}\Phi_0} &=   \frac{f_1(x_1,t_1)}{f_0(x_0,t_1)} \frac{\alpha_\mathrm{S}(t_1)}{\alpha_\mathrm{S}(\mu_\mathrm{R})} \frac{f_2(x_2,t_2)}{f_1(x_1,t_2)}  \frac{\alpha_\mathrm{S}(t_2)}{\alpha_\mathrm{S}(\mu_\mathrm{R})} \Pi_{\mathcal{S}_0}(t_0, t_1) \Pi_{\mathcal{S}_1}(t_1, t_2) \, \mathrm{B}_2 \equiv w'_2 \mathrm{B}_2,
\end{align}
where $w_n$ differs from $w'_n$ only by a factor of $\Pi_{\mathcal{S}_n}(t_n, t_\mathrm{MS})$. The sum of these cross sections is the $+2$-jet inclusive CKKW-L merged cross section, 
\begin{equation}
\label{eq2.18:CKKWLsigma2jet}
     \sigma_2^\mathrm{CKKW-L}  = \int \mathrm{d}\Phi_0 \int \mathrm{d}t \Bigg( w_0 \mathrm{B}_0 + \int \mathrm{d}t_1 w_1 \mathrm{B}_1 + \int \int \mathrm{d}t_1 \mathrm{d}t_2 w'_2 \mathrm{B}_2 \Bigg)
\end{equation}
which can reproduce the Born-level cross section only if certain conditions are met, one being that the parton-shower expansion exactly reproduces the matrix element contributions for the first two emissions. Further conditions arise for multiplicities with $n>1$, for example identical phase space coverage and no-branching probabilities. The task of generating PS emissions according to the ME is possible to be done process-by-process, and has potential for a future study on \pythia DIS improvements. These so called matrix-element corrections have been implemented for various parton shower algorithms and processes \cite{Giele:2007di,Bengtsson:1986hr,Fischer:2017yja}. 

Considerable work has been done to preserve the parton-shower unitarity in multi-jet merging, and can be achieved using the UMEPS algorithm \cite{L_nnblad_2013,Platzer:2012bs,Bellm:2017ktr}. The UMEPS scheme, which stands for Unitary Matrix Element + Parton Shower, restores the shower unitarity by integrating over multi-jet states and inducing resummation in lower multiplicity states. This step is straightforward in the CKKW-L algorithm during the construction of PS histories.
Following again \cite{L_nnblad_2013}, let us use the following notation for these integrated states and the UMEPS weight
\begin{align}
    \frac{\mathrm{d}\sigma_n^\mathrm{in}}{\mathrm{d}\Phi_0} &= w'_n \mathrm{B}_n \equiv \widehat{\mathrm{B}}_n, & \frac{\mathrm{d}\sigma_{n\rightarrow m}^\mathrm{in}}{\mathrm{d}\Phi_0} = \int \mathrm{d}^{n-m}\Phi \, w'_n \mathrm{B}_n \equiv \int \widehat{\mathrm{B}}_{n\rightarrow m},
\end{align}
so that the UMEPS cross section can be written as a sum of positive-weight $n$-jet samples and negative-weight $(n-m)$-jet samples. Here the expression $\int \mathrm{d}^{n-m}\Phi \, w'_n \mathrm{B}_n$ contains integrations of states $\mathcal{S}_{n-1}, \mathcal{S}_{n-2}, ..., \mathcal{S}_{m+2}, \mathcal{S}_{m+1}$. In this way it is possible to induce resummation in $m$-jet cross sections by integrating over the ($n > m$)-jet states $n-m$ times. In practice, this integration is the replacement of state $\mathcal{S}_n$ with state $\mathcal{S}_{n-1}$ while keeping the weight of the original state. These states are already obtained while constructing PS histories in the CKKW-L procedure. In the case of $+2$-jet merging, the inclusive cross section becomes 
\begin{equation}
    \sigma_2^\mathrm{UMEPS} = \int \mathrm{d}\Phi_0 \Bigg[ \Bigg( \widehat{\mathrm{B}}_0 - \int \widehat{\mathrm{B}}_{1\rightarrow 0} - \int \widehat{\mathrm{B}}_{2\rightarrow 0} \Bigg) 
    + \Bigg( \int \widehat{\mathrm{B}}_1 - \int \widehat{\mathrm{B}}_{2\rightarrow 1} \Bigg) 
    + \int \widehat{\mathrm{B}}_2 \Bigg].
\end{equation}
This add-and-subtract approach captures the relevant effects of multi-jet merging while retaining the Born-level inclusive cross section. 
It should be noted again that the number of shower histories is minimal when using the above algorithm with \vincia's sector-merging approach. That is, there is only one history when only leading-colour gluon emissions are present and the number of histories scales at most factorially with the number of same-flavour quark pairs when also gluon splittings are present.

\subsection{Deep inelastic scattering}
\label{sec2.3:dis}
High-energy electron-proton collisions have been studied extensively in the past in order to obtain information about the structure of the proton and its scale dependence. The electron as a point-like probe is a clean candidate for studies of parton distribution functions, but enables also extraction of information about QCD jets. In these scattering events, the interaction between the electron and a parton is mediated by a highly virtual photon breaking the proton apart, which is why the process is considered as deeply inelastic. The proton is broken up into its constituent parts, which gain a significant amount of energy in the collision. The scattering of electrons and protons is characterized by the following Lorentz-invariant kinematical quantities
\begin{align}
    x &= \frac{-q^2}{2(p \cdot q)} \text{ Bjorken $x$,}  \\
    Q^2 &= -q^2 = -(k-k')^2 \text{ intermediate photon virtuality,} \\
    y &= \frac{q \cdot p}{k \cdot p} \text{ inelasticity, the lepton energy-loss fraction in proton rest frame, and } \\
    W^2 &= (p+q)^2 \text{ invariant mass of the hadronic system.}
\end{align}

\begin{figure}
\centering
 \begin{minipage}[c]{0.4\textwidth} 
    \includegraphics[]{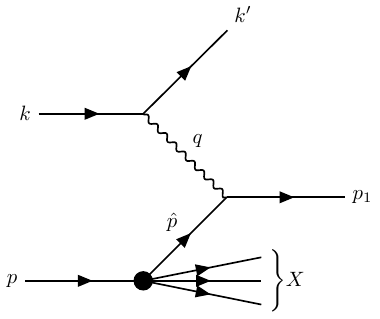}
  \end{minipage}\hspace{2em}
  \begin{minipage}[c]{0.4\textwidth}
    \includegraphics[]{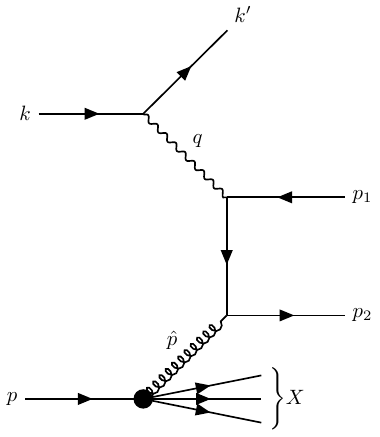}
  \end{minipage}%
        
    \caption{Feynman diagrams of deep inelastic scattering in the parton model. The incident lepton has four-momentum $k$, proton $p$, and quark $\hat{p}$. The lepton and quark interact via a photon with four-momentum $q$, resulting in outgoing lepton with $k'$, quark with $\hat{p}$ and the proton remnants $X$. The left diagram is the Born level process $e^- p \rightarrow e^- j$, whereas the right diagram is one of the possible diagrams of $e^- p \rightarrow e^- 2j$. }
    \label{fig2:PartonlevelFeynman}
\end{figure}

The left diagram of figure~\ref{fig2:PartonlevelFeynman} is the Born-level process, corresponding to the lowest multiplicity state $\mathcal{S}_0$ with no QCD interactions present. It represents the graph of the cross section $\mathrm{B}_0$ of eq.~\ref{eq2.11:sigmaPS}. This process is of order $\alpha_\mathrm{EM}^2$. The process on the right side is $\gamma^* g \rightarrow q \Bar{q}$, one of the processes contributing to the cross section $\mathrm{B}_1$, with order $\alpha_\mathrm{EM}^2 \alpha_\mathrm{S}$. There are other processes belonging to the same state $\mathcal{S}_1$, for example $\gamma^* q \rightarrow q g$. The outgoing partons eventually hadronize and are detected as jets of particles, with momentum proportional to the initial outgoing parton. 
An important frame choice in the context of DIS is the Breit frame, also known as the infinite-momentum frame. In this frame, the proton is boosted to a very high velocity in the $z$-direction, and the virtual photon four-vector only has a spatial component $q_z = -Q$. The frame can be defined at parton level by requiring 
\begin{equation}
    \hat{p}^2 = 2 \xi p \cdot q + q^2 = 0 \, \Leftrightarrow \, \xi = \frac{Q^2}{2(p\cdot q)}, 
\end{equation}
where $\xi$ is the fraction of proton's three-momentum carried by the parton $\hat{p}$. It is evident that in this frame, the variable $x$ is equivalent to $\xi$, and the parton momenta can be expressed as $\hat{p} = xp$. A notable property of this frame is that there is no transverse momentum involved in the Born-level process: particles can only acquire transverse momentum in higher final-state multiplicities.  

\subsection{Merging in DIS}
\label{sec2.4:DISmerging}
Scattering processes in DIS might involve multiple hard scales in the same event. Some of the effects from low-virtuality, high jet transverse-momentum might be lost on the simplest choice of $\mu_\mathrm{F}^2 = Q^2$, so other scale choices have been studied, for example taking the dijet transverse momenta into account with $\mu_\mathrm{F}^2 = (Q^2 + \langle p_\mathrm{T}^2\rangle) / 2$, in ref.~\cite{Currie_2017}. The choice for the factorization scale has crucial implications in the context of merging, since this scale is used as the shower starting scale. For example the process depicted in the right diagram of figure \ref{fig2:PartonlevelFeynman} has the possibility of having high transverse momenta jets $p_{\mathrm{T},i}^2 > Q^2$. A logical choice is to choose the factorization and renormalization scales as a combination of virtuality $Q^2$ and the Breit frame jet $p_\mathrm{T}$'s. 

A fixed merging scale is sufficient for processes that can be characterised by a single scale, such as Drell-Yan or electron-positron-annihilation where the virtuality of the intermediate boson serves as a characteristic scale. This approach is insufficient whenever multiple scales are present, such as in DIS where the transverse momentum in the Breit frame and the virtuality of the exchanged boson characterize intrinsically different phase-space regions. 
Notably, VBF-processes, sometimes viewed as ``double-DIS'' processes \cite{Han:1992hr}, are sufficiently well described by scale choices considering the transverse masses of the Boson and Born-level jets \cite{Buckley:2021gfw,Hoche:2021mkv}.

We denote $\Tilde{t}_\mathrm{MS}$ as the fixed-value merging scale which separates low-transverse-momenta jets from hard jets. It serves as the cut in the merging algorithm, in principle assigning vetoes to matrix-element events below this scale. As discussed above, the hardest scale in a DIS event is not unique, and this handling with the usual choice of $\mu_\mathrm{F}^2 = Q^2$ might miss a large portion of the low-$Q^2$, high-$p_\mathrm{T}$ phase space. A re-definition of the merging scale aims to provide a similar effect as switching to a more suitable factorization scale. To account for the contributions of these processes, a dynamic merging scale is defined as in the study \cite{Carli_2010}, as shown by the following equation: 
\begin{equation}
    \label{eq2.26:DynamicTMS}
    t_\mathrm{MS} = \frac{\Tilde{t}_\mathrm{MS}}{1+ \frac{\Tilde{t}_\mathrm{MS}}{Q^2 S^2}},
\end{equation}
where $\Tilde{t}_\mathrm{MS}$ is the fixed merging scale value and $0 < S \leq 1$ should be chosen so that the matrix-element domain covers the low-$Q^2$, high-$p_\mathrm{T}$ phase space. The dynamic merging scale simply lowers the cut value, especially in the low-$Q^2$ region, allowing the matrix element sample to handle a larger portion of this phase space.

\section{Computational setup and validation}
\label{sec3:setup}
In this section, we review the key computational choices. For the generation of the parton-level events, this includes renormalization and factorization scales, PDFs, and electroweak parameters. For the merging implementation we discuss the algorithms used and the corresponding merging scales. Additionally, we conduct preliminary analyses to validate three different setups with appropriate algorithm and scale choices. 

\subsection{Computational setup}

Hard events are generated at leading order in QCD using a HPC-enabled version of the \sherpa event generator \cite{Gleisberg:2008ta}, using the automated matrix-element generator COMIX \cite{Gleisberg_2008} and stored using the efficient and scalable HDF5 event-file format \cite{Hoche:2019flt,Bothmann:2023ozs}.\footnote{Available at \url{https://gitlab.com/hpcgen/me}.} 
For cross checks, we have also generated events with \madgraph \cite{Alwall_2014} using identical model parameters and phase-space cuts as the \sherpa samples, stored using the Les Houches Event File format \cite{Alwall:2006yp}. 
All parton-multiplicity samples are generated with $10~\mathrm{M}$ events, except the highest multiplicity sample, which has $1~\mathrm{M}$ events of the process $e^- ~p \rightarrow e^- ~5j$. We consider a larger event sample unreasonable for the scope of this study, as the generation of unweighted events requires a significant investment of CPU resources (approximately $24~\mathrm{k}~\mathrm{CPUh}/1~\mathrm{M}~\text{events}$). With \madgraph we limit our study to events with up to $2$ additional partons, as we observed that the cross section became increasingly unstable with $3$ additional partons or more with the applied cuts. 
We have made all event samples publicly available \cite{Sample1,Sample2,Sample3,Sample4,Sample5}. 
The cuts are chosen very inclusively to cover multiple analyses and kinematic ranges with the cost of lower statistics in some multi-differential observables. The lowest-multiplicity sample is generated with a cut on virtuality $Q^2 > 2$ GeV$^2$, while for higher multiplicities we further employ the longitudinally invariant inclusive $k_\mathrm{T}$ algorithm in the FastJet \cite{Cacciari_2012} implementation, in both the analyses and event generators, and apply an additional cut of $k_\mathrm{T} > 2~\mathrm{GeV}$ in the center-of-momentum frame of the beams to regulate infrared divergences. 

We use the $G_\upmu$-scheme with electroweak input parameters
\begin{equation}
    m_W = 80.419~\mathrm{GeV} \, , \quad m_Z = 91.188~\mathrm{GeV} \, , \quad G_\mathrm{F} = 1.16639\times 10^{-5}~\mathrm{GeV}^{-2} \, .
\end{equation}
Quark masses are taken in the four-flavour scheme and the $b$-quark mass is set to $m_b = 4.7~\mathrm{GeV}$.
The PDF choice is \texttt{NNPDF40\_lo\_pch\_as\_01180} \cite{NNPDF:2021njg} interfaced via the LHAPDF6 package \cite{Buckley_2015}. According to the PDF, we use a value of $\alpha_\mathrm{S}(m_Z) = 0.118$ in the hard scattering.

We consider two different renormalization- and factorization-scale choices, namely
\begin{eqnarray}
    \label{eq3.2:muF_Q2}
    1.) && \mu_\mathrm{R}^2 = \mu_\mathrm{F}^2 = Q^2 \, , \\
    \label{eq3.3:muF_Q2+pT2}
    2.) && \mu_\mathrm{R}^2 = \mu_\mathrm{F}^2 = \frac{1}{2}\left(Q^2 + H_{\mathrm{T}}^2\right) \, .
\end{eqnarray}
and their variations by a factor of $2$, which is a standard way of quantifying the uncertainty of theory calculations and Monte Carlo predictions. 
Here, $p_{\mathrm{T}}$ refers to the transverse momentum in the Breit frame and its scalar sum is taken as
\begin{equation}
\label{eq3.4:HT}
    H_{\mathrm{T}} = \sum\limits_{i=1}^N p_{\mathrm{T},i} \, \, \text{, for }N \text{-parton final states.} 
\end{equation}%
This scale choice takes into account events for which $Q^2$ is not necessarily the hardest energy scale of the process. Notice that in case of only one outgoing parton $H_\mathrm{T} = 0$ in the Breit frame. When using a dynamic merging scale of eq. \eqref{eq2.26:DynamicTMS}, our default choice of the reference scale is $\tilde{t}_\mathrm{MS} = (5~\mathrm{GeV})^2$ and $S=0.7$ for the $S$-parameter, which we have observed to yield good overall agreement with data. The dynamic merging scale is to be used with the standard choice of factorization scale $\mu_\mathrm{F}^2 = Q^2$. 

Reflecting the above discussion, we consider three separate merging setups: 
\begin{itemize}
    \item CKKW-L with dynamic merging scale as in eq. \eqref{eq2.26:DynamicTMS} and $ \mu_\mathrm{R}^2 = \mu_\mathrm{F}^2 = Q^2$,
    \item CKKW-L with fixed merging scale and $\mu_\mathrm{R}^2 = \mu_\mathrm{F}^2 = \frac{1}{2} (Q^2 + H_\mathrm{T}^2)$ as in eq. \eqref{eq3.3:muF_Q2+pT2},
    \item UMEPS with fixed merging scale and $\mu_\mathrm{R}^2 = \mu_\mathrm{F}^2 = \frac{1}{2} (Q^2 + H_\mathrm{T}^2)$ as in eq. \eqref{eq3.3:muF_Q2+pT2}.
\end{itemize}
The scale choice of eq. \eqref{eq3.3:muF_Q2+pT2} and the dynamical merging scale both contribute to the low-virtuality phase space, and using both options led to too much of the desired effect. A setup with only one of these options lead to most realistic modelling in our studies, thus we only consider the three options listed above. 

Parton-level events are merged with \pythia 8.3, using the \vincia antenna shower \cite{Brooks:2020upa} in conjunction with its dedicated sectorized multi-jet merging framework \cite{Brooks:2020mab}. The necessary modifications to enable merging in DIS are included in \pythia version 8.313 onwards. The hadronization and fragmentation of showered configurations is performed with \pythia's Lund string-fragmentation model \cite{Andersson:1983ia}, using \vincia's preliminary default tune \cite{Brooks:2020upa}, which has been derived based on the methods developed for \pythia's Monash 2013 tune \cite{Skands:2014pea}. 
In the shower evolution, we use a two-loop running of the strong coupling with $\alpha_\mathrm{S}(m_Z)=0.118$ in the $\overline{\mathrm{MS}}$ scheme. The effective coupling is evaluated in the CMW scheme \cite{Catani:1990rr} with additional renormalization-scale factors \cite{Brooks_2020}.
Analyses are performed using \pythia's internal analysis routines and the \rivet framework \cite{Bierlich:2019rhm}, for which we have prepared new analyses for HERA data. The details of these analyses are discussed in section~\ref{sec4:results}. 

\begin{figure}
    \centering
    \includegraphics[scale=0.6]{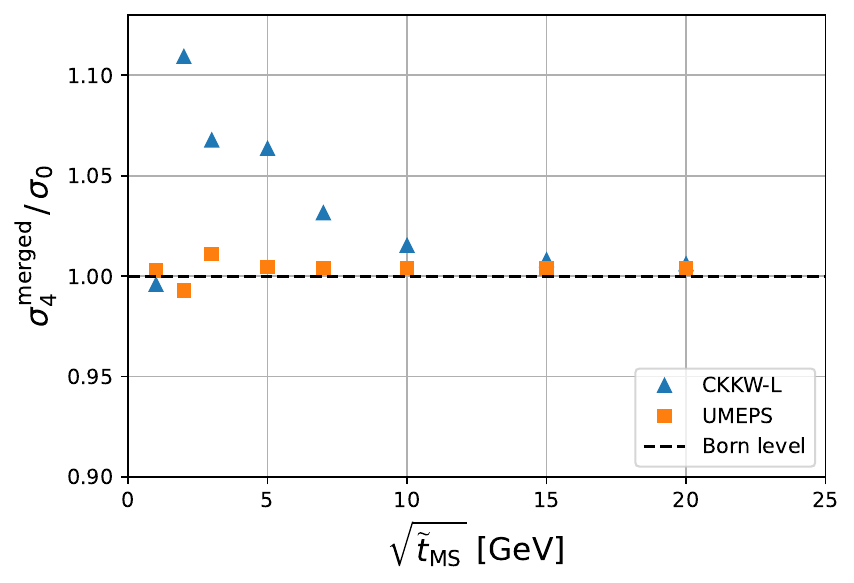}
    \caption{Merged cross sections $\sigma^\mathrm{merged}_4$ of CKKW-L (red) and UMEPS (green) algorithms in $+4$-jet merging as ratios to the Born level cross section $\sigma_0$. Both algorithms use the fixed merging scale and $\mu_\mathrm{F}^2 = \frac{1}{2}\left(Q^2 + H_\mathrm{T}^2\right)$. Cross section ratios are presented as a function of merging scale $\Tilde{t}_\mathrm{MS}$. }
    \label{fig3:sigma_tms_plot}
\end{figure}

\subsection{Validation}

Before comparing to experimental data we demonstrate the key features of the merging algorithms and the effects of different setups. Figure~\ref{fig3:sigma_tms_plot} includes full merged cross sections from runs with different values for the merging scale parameter $\Tilde{t}_\mathrm{MS}$. Results are calculated with CKKW-L and UMEPS algorithms, using a fixed merging scale in both cases. The cross sections are plotted as ratios of the lowest multiplicity leading-order cross section $\sigma_0$. The figure indicates that CKKW-L violates the unitarity of the parton shower, especially with lower values of the merging scale, where differences are of the order of $10\%$. The UMEPS-merging consistently achieves a value similar to $\sigma_0$ regardless of the merging scale. 

\begin{figure}
    \centering
    \hspace{-3em}
        \begin{minipage}{.295\textwidth}
            \includegraphics[scale=0.45]{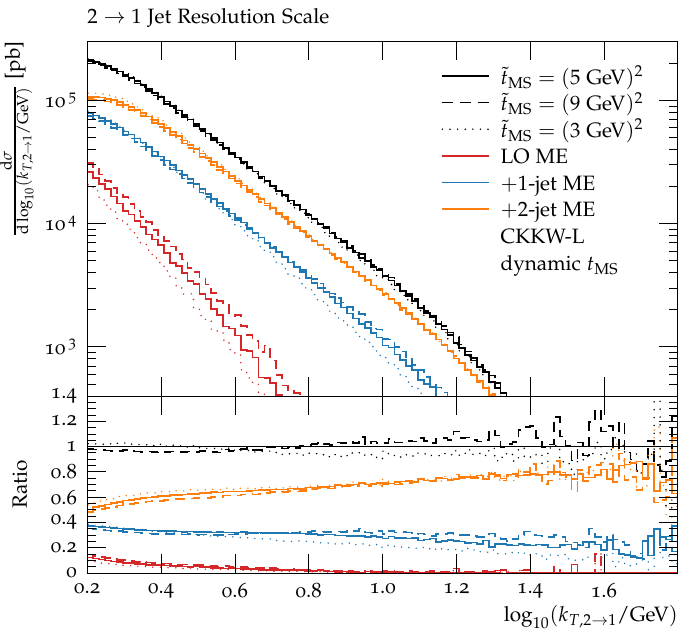}
        \end{minipage}
        \begin{minipage}{.295\textwidth}
            \includegraphics[scale=0.45]{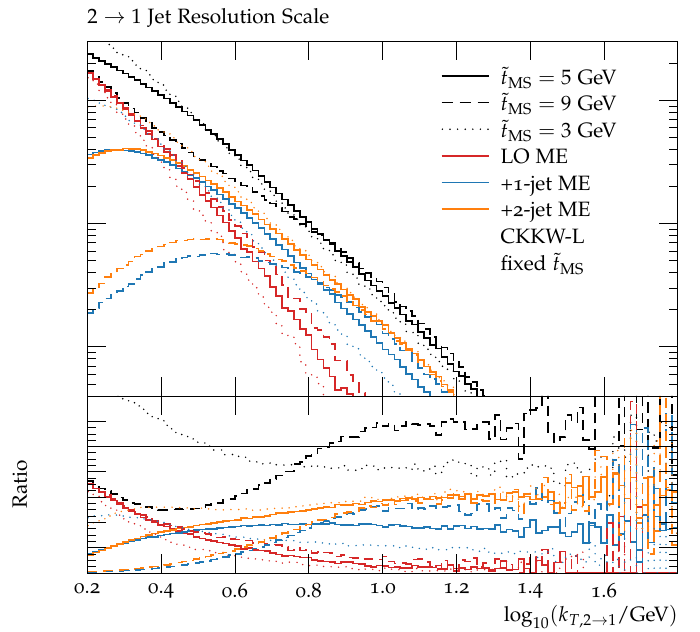}
        \end{minipage}
        \begin{minipage}{.295\textwidth} 
            \includegraphics[scale=0.45]{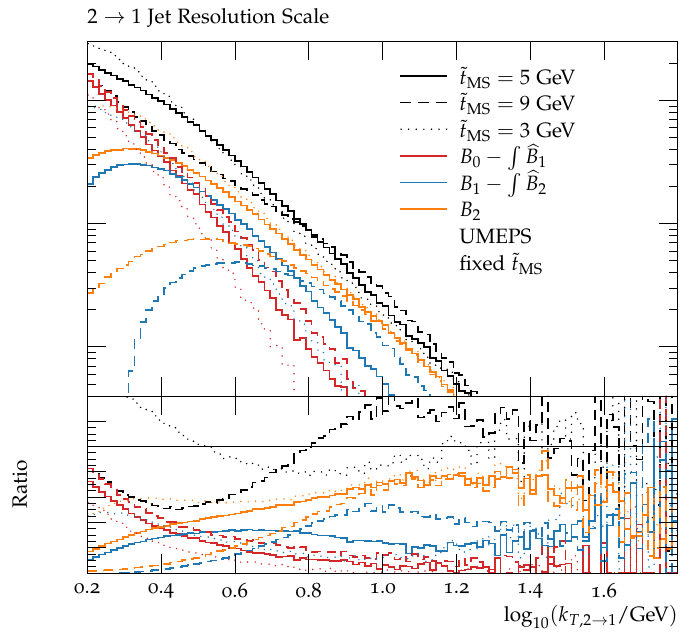}
        \end{minipage}
    \caption{Jet rates with variation in $\Tilde{t}_\mathrm{MS}$, using CKKW-L  with a dynamic merging scale and $\mu_\mathrm{F}^2 = Q^2$ (left) and CKKW-L (middle) and UMEPS with a fixed merging scale and $\mu_\mathrm{F}^2 = \frac{1}{2}\left(Q^2 + H_\mathrm{T}^2 \right)$ (right) with $+2$ jets. The full merged results (black line) are the sums of contributions from the jet multiplicity samples (colored lines), with variations in the parameter $\Tilde{t}_\mathrm{MS}$ shown as dashed and dotted lines. }
    \label{fig4:TMSvariation}
\end{figure}

\begin{figure}
    \centering
    \begin{minipage}{.49\textwidth}
        \includegraphics[scale=0.4]{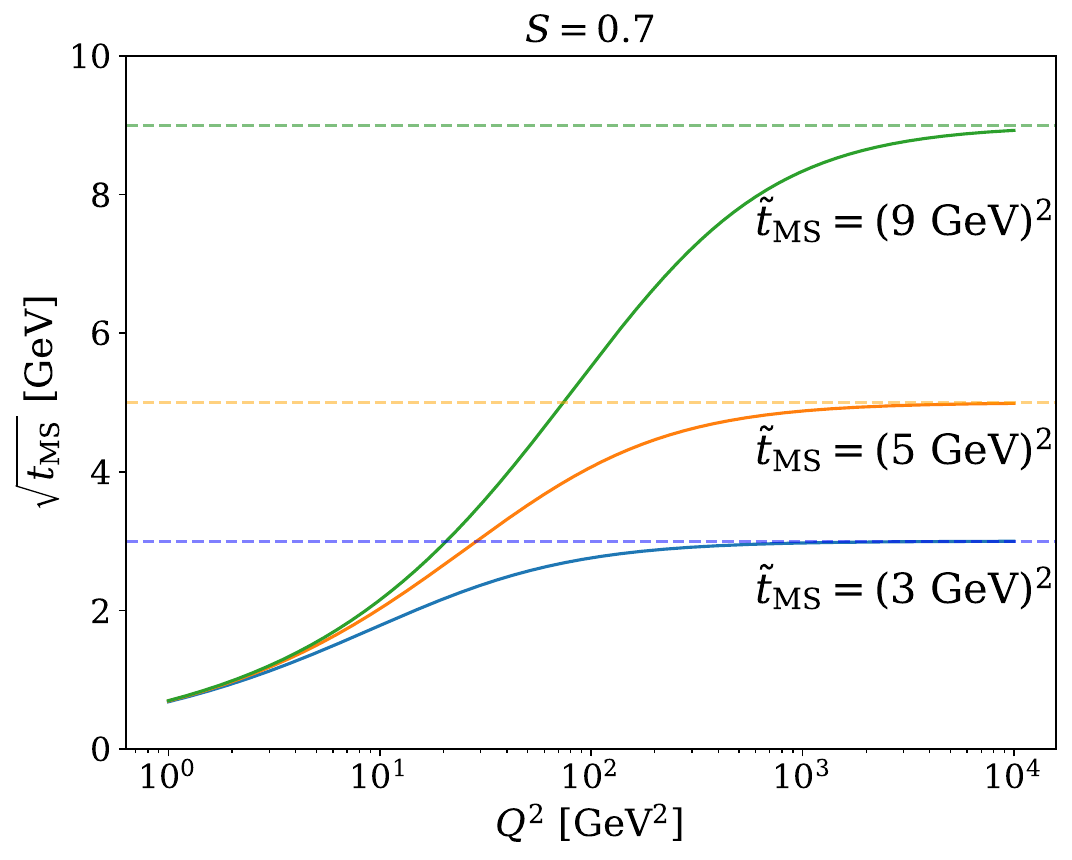}
    \end{minipage}
    \begin{minipage}{.49\textwidth}
        \includegraphics[scale=0.4]{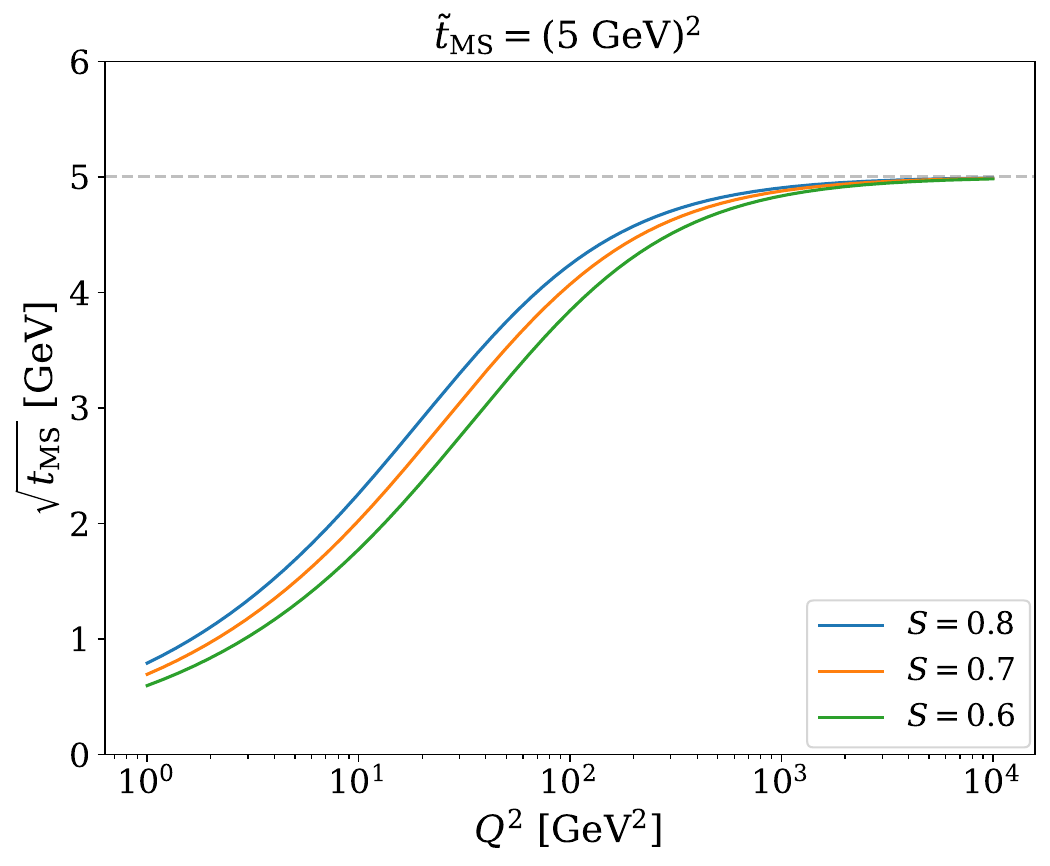}
    \end{minipage}
    \caption{Dynamic merging scale $t_\mathrm{MS}$ as a function of $Q^2$ with different values of parameters $\Tilde{t}_\mathrm{MS}$ (left) and $S$ (right). }
    \label{fig5:DynamicMergingScale}
\end{figure}

The exact definition of a merging scale should not affect the full inclusive merged cross section too much, as it dictates the ratios of different contributions of the full cross section. Just as is the case for factorization and renormalization scales, the final cross sections should ideally not be too sensitive to different scale choices. New parameters are introduced in merging algorithms, and the effects of varying these parameters should also be a part of the uncertainty analysis in multi-jet merging predictions. 
The effect of varying the fixed value of $\Tilde{t}_\mathrm{MS}$ is shown in figure \ref{fig4:TMSvariation} for the three considered setups. The $2 \rightarrow 1$ resolution scale (or differential jet rate), $k_{\mathrm{T},2\rightarrow 1}$, is the transverse momentum scale at which the event transitions from a two-jet state to a one-jet state. It is related to the value of a parton-shower evolution parameter at a 1-to-2 splitting. The full results from each setup (black lines) are the sums of contributions from each event sample, with the $+0$-jet $(e^- ~p \rightarrow e^- ~j)$ sample in red, $+1$-jet $(e^- ~p \rightarrow e^- ~2j)$ in blue and $+2$-jet $(e^- ~p \rightarrow e^- ~3j)$ in orange. Variations of the merging-scale parameter are plotted as dashed lines for $(9~\mathrm{GeV})^2$, dotted lines for $(3~\mathrm{GeV})^2$ and the central value $(5~\mathrm{GeV})^2$ as solid lines. Figure \ref{fig4:TMSvariation} shows how similar results are obtained with different setups, and how the value of the merging scale affects each setup. We find that while contributions from each multiplicity are sensitive to the $\Tilde{t}_\mathrm{MS}$ choice, especially in case of a fixed merging scale, the total cross section is always less affected. The contributions $\widehat{\mathrm{B}}_n - \int \widehat{\mathrm{B}}_{n+1 \rightarrow n} - ... - \int \widehat{\mathrm{B}}_{n+1 \rightarrow 0}$ for UMEPS merging can be compared to contributions $\mathrm{B}_n$ of CKKW-L merging. With a dynamic merging scale, small-$Q^2$ dynamics are less sensitive to $\Tilde{t}_\mathrm{MS}$, while the fixed merging results vary considerably more. This can be explained with the weak sensitivity of $t_\mathrm{MS}$
to $\Tilde{t}_\mathrm{MS}$ at low-$Q^2$ region as shown in figure \ref{fig5:DynamicMergingScale}. Using a fixed merging scale results to a high variation across jet rates in lower values of $k_\mathrm{T}$. This effect holds even when restoring unitarity with the UMEPS algorithm. However, the experimental data confronted in section~\ref{sec4:results} have cuts on jet transverse momenta $p_\mathrm{T} \sim 4$ GeV, which roughly translates to $\log_{10}(k_\mathrm{T}/\mathrm{GeV}) \sim 0.6$, so the high variation in low-$k_\mathrm{T}$ region is not present in the final results. Significantly reduced variations of the individual contributions in case of dynamical merging-scale determination makes it clearly an appropriate choice for multi-jet merging in DIS. The combined jet rates are, however, still fairly similar in all of the studied cases and thus also the algorithms with fixed merging scale can be considered as applicable options.

\section{Results}
\label{sec4:results}

Using parton-level events with up to five final-state partons from \sherpa, combined with \vincia's merging machinery in \pythia, we have simulated full hadron-level DIS events. The results after merging events of different multiplicities and parton showers are compared to data from HERA experiments. We have produced two new \rivet jet analyses, the first one being \textbf{H1\_2002\_I588263} \cite{H1:2002qhb} with data collected from experiments conducted in 1996 and 1997 with beam energies of $E_p = 820$ GeV and $E_e = 27.5$ GeV and experimental cuts on virtuality $5.5 < Q^2 < 80 ~\mathrm{GeV}^2$ and inelasticity $0.2 < y < 0.6$ as well as cuts on Breit-frame jet transverse momentum $p^\mathrm{jet}_\mathrm{T} > 4 ~\mathrm{GeV}$ and laboratory-frame jet pseudorapidity $-1 < \eta < 2.5$. The second analysis is \textbf{H1\_2016\_I1496981} \cite{H1:2016goa} with data from 2005 to 2007 and higher energies of $E_p = 920$ GeV and $E_e = 27.6$ GeV, with similar cuts of $5 < Q^2 < 100 ~\mathrm{GeV}^2$, $0.2 < y < 0.6$, Breit-frame jet $E_\mathrm{T} > 5 ~\mathrm{GeV}$ and laboratory-frame $-1 < \eta < 2.8$. Note that in both analyses the jet transverse momentum (energy) is measured in the Breit frame, while the jet pseudorapidity is measured in the laboratory frame. Both analyses will be included as public routines in a future release of \rivet and all results produced in this study are supplemented to the electronic version of this paper. We will mainly focus on the more recent H1 analysis as, in addition to inclusive jet measurements, it also provides differential cross sections for di- and trijets. There are jet measurements available also from the ZEUS collaboration, e.g. ref.~\cite{ZEUS:2010vyw}, but these are at larger values of $Q^2$ where the impact of multi-jet merging is reduced \cite{Carli_2010} and thus not considered in this study. 

\begin{figure}
    \centering
    \hspace{-3em}
    \begin{minipage}{.295\textwidth}
        \includegraphics[scale=0.45]{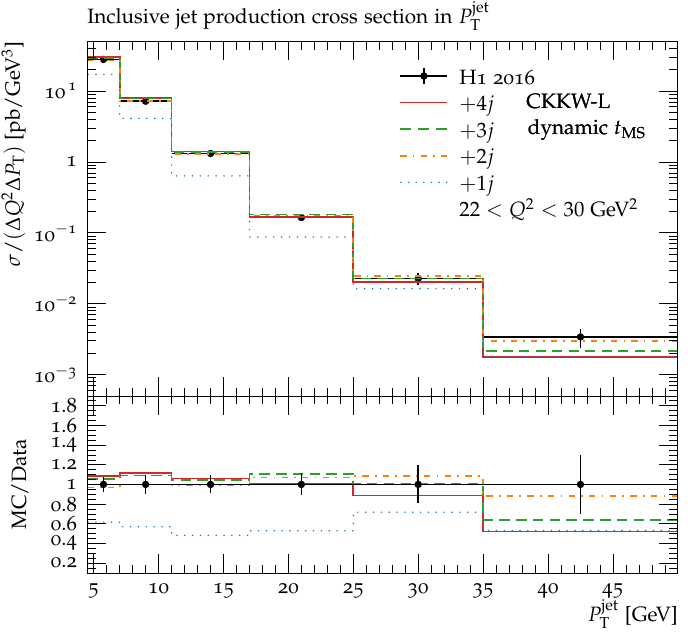}
    \end{minipage}
    \begin{minipage}{.295\textwidth}
        \includegraphics[scale=0.45]{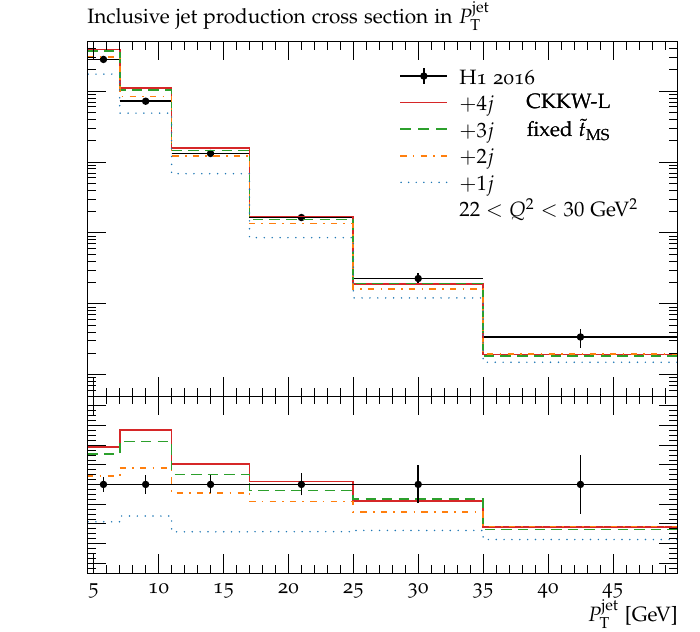}
    \end{minipage}
    \begin{minipage}{.295\textwidth}
        \includegraphics[scale=0.45]{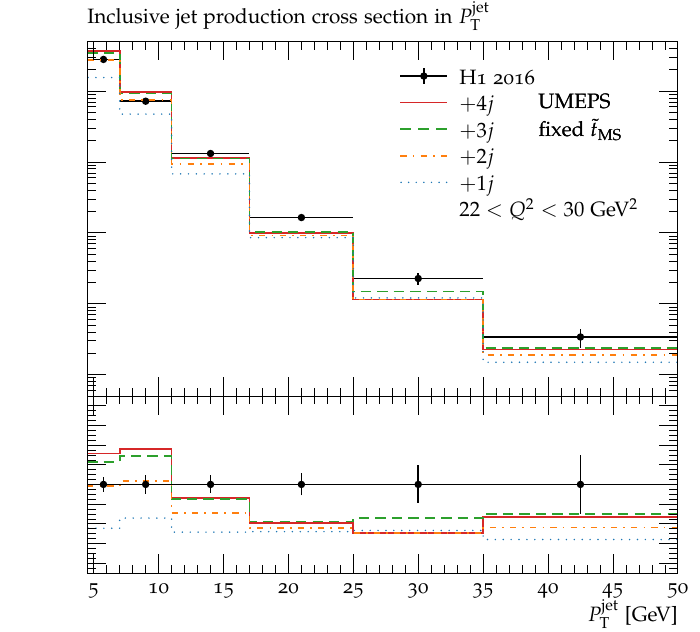}    \end{minipage}
    \caption{Jet production cross sections in terms of inclusive jet transverse momenta in the range $22 < Q^2 < 30$ GeV$^2$. Different amount of additional final state partons are shown in the three merging setups: dynamic CKKW-L merging with $\mu_\mathrm{F}^2 = Q^2$ and fixed CKKW-L and UMEPS merging with $\mu_\mathrm{F}^2 = \frac{1}{2}\left(Q^2 + H_\mathrm{T}^2 \right)$. Data from H1 \cite{H1:2016goa}. }
    \label{fig6:progression_incjets}
\end{figure}
\begin{figure}
    \centering
    \hspace{-3em}
    \begin{minipage}{.295\textwidth}
        \includegraphics[scale=0.45]{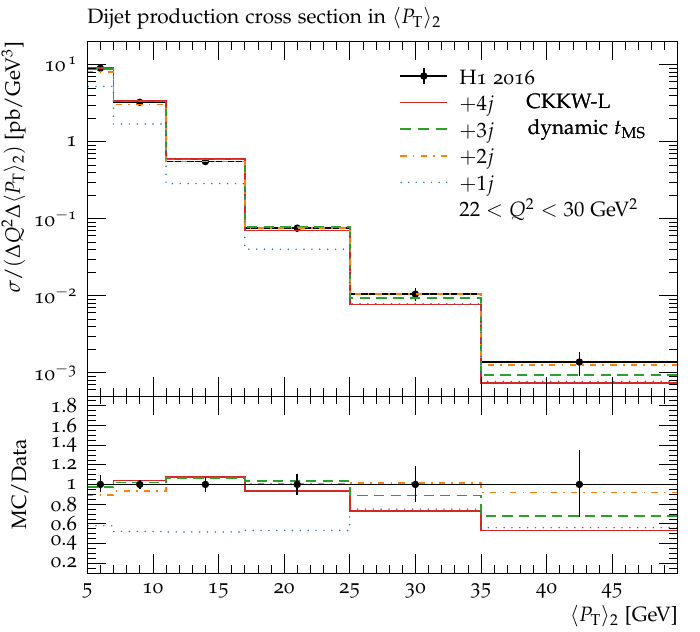}
    \end{minipage}
    \begin{minipage}{.295\textwidth}
        \includegraphics[scale=0.45]{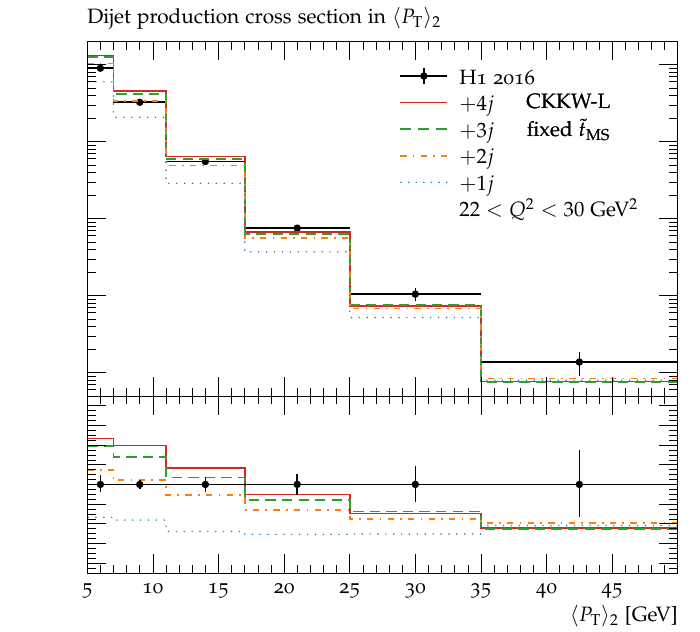}
    \end{minipage}
    \begin{minipage}{.295\textwidth}
        \includegraphics[scale=0.45]{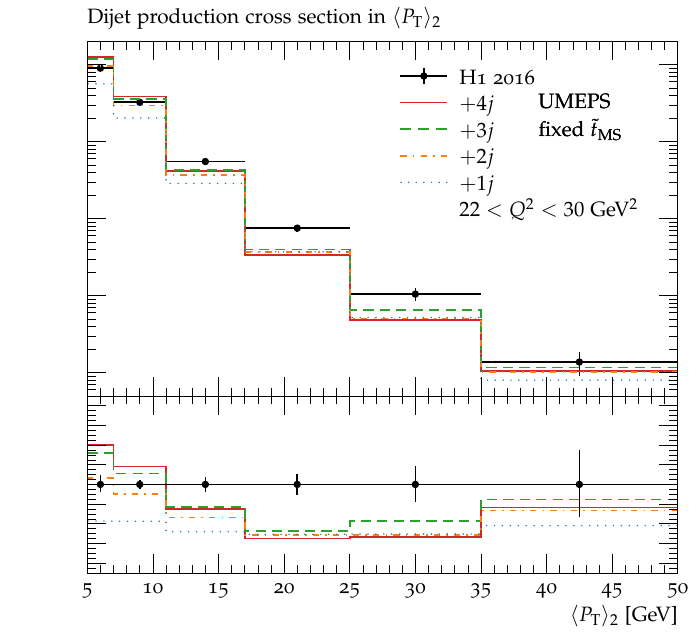}
    \end{minipage}
    \caption{Dijet production cross sections in terms of average of transverse momenta of two hardest jets in the range $22 < Q^2 < 30$ GeV$^2$. Data from H1 \cite{H1:2016goa}. Details of the plot are explained in figure \ref{fig6:progression_incjets}. }
    \label{fig7:progression_dijets}
\end{figure}
\begin{figure}
    \centering
    \hspace{-3em}
    \begin{minipage}{.295\textwidth}
        \includegraphics[scale=0.45]{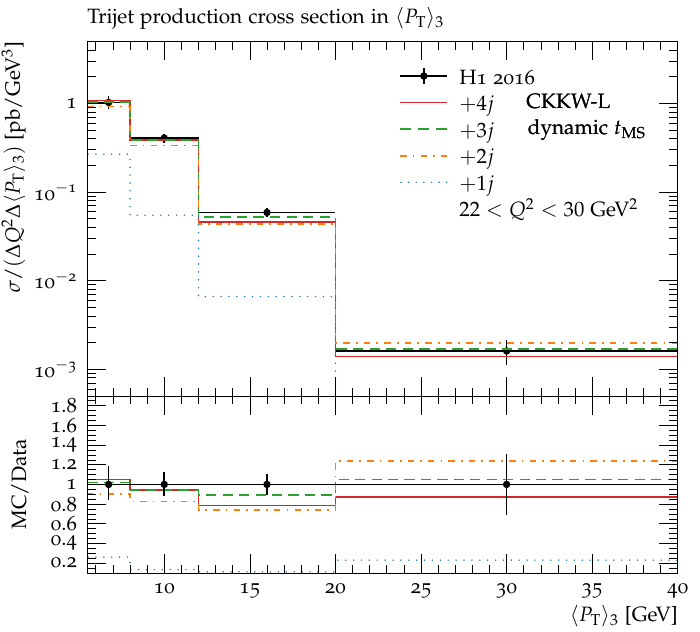}
    \end{minipage}
    \begin{minipage}{.295\textwidth}
        \includegraphics[scale=0.45]{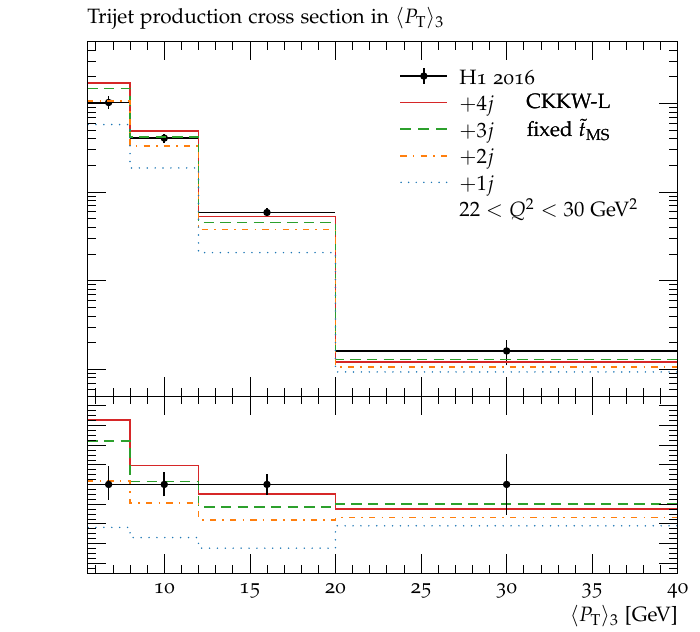}
    \end{minipage}
    \begin{minipage}{.295\textwidth}
        \includegraphics[scale=0.45]{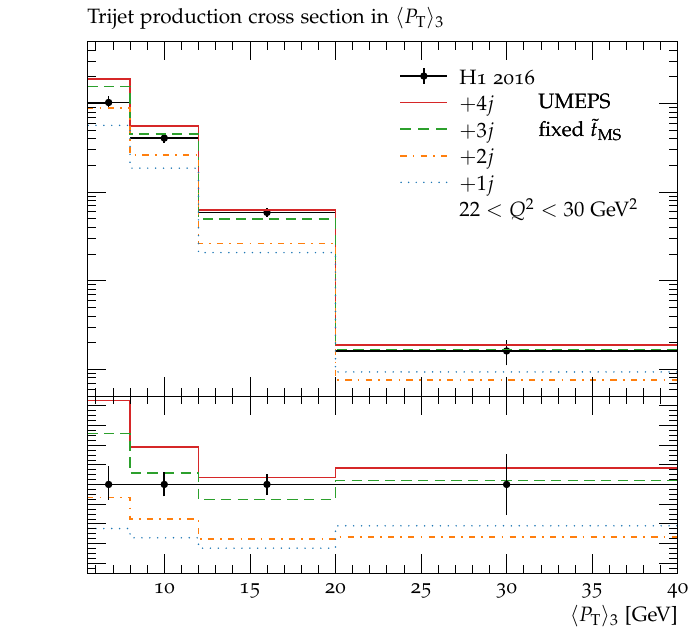}
    \end{minipage}
    \caption{Trijet production cross sections in terms of average of transverse momenta of three hardest jets in the range $22 < Q^2 < 30$ GeV$^2$. Data from H1 \cite{H1:2016goa}. Details of the plot are explained in figure \ref{fig6:progression_incjets}. }
    \label{fig8:progression_trijets}
\end{figure}

\subsection{Jet multiplicities}
\label{sec4.1:JetMultiplicities}

With the dynamic merging-scale prescription in CKKW-L, discussed in section~\ref{sec2.4:DISmerging}, the event generation produces final states with multiple jets which generally originate from the additional hard partons in the event samples. The choice of a dynamic merging scale aims to ensure that hard partons are more often generated by the matrix-element generator with decreasing $Q^2$. The UMEPS scheme also produces multi-jet final states, but yields both positive and negative weight samples as discussed in \ref{sec2.2:merging}, which retain the Born-level cross section when additive and subtractive contributions of different multiplicities are combined. 

We extend these merging setups up to $+4$-jet merging, and compare the results to H1 data in figures \ref{fig6:progression_incjets}, \ref{fig7:progression_dijets} and \ref{fig8:progression_trijets}. Here we only consider a $Q^2$ range of $22 < Q^2 < 30 ~\mathrm{GeV}^2$ as an example and study the $Q^2$ dependence in subsection~\ref{sec4.2:MergingScales}. The dashed lines represent full merged results with the lower multiplicity samples, displaying how the jet transverse momentum distributions converge as more and more final-state partons are included in the core process. The analysis accepts all jets with $k_\mathrm{T}\geq 4$ GeV, reconstructed with the FastJet $k_\mathrm{T}$-algorithm \cite{Cacciari_2012} and a cone size of $R=1.0$. 
The inclusive jet-production cross section in figure \ref{fig6:progression_incjets} accumulates all jets in the fiducial region, meaning that a single event might contribute multiple times to the same histogram.
With $+1$-jet merging, the description is not sufficient, and the simulation clearly misses some of the spectrum compared to the data. For inclusive jet $p_\mathrm{T}$ distributions, $+2$-jet merging still improves description of the data significantly, but adding further samples to the merging leads to converging distributions close to $+2$-jet results. This reflects the fact that the value of a jet cross section decreases with increasing jet multiplicity: even though the $+4$-jet sample has at least five jets per event on average, these events are much more unlikely, i.e. have a smaller cross section than the lower-multiplicity events. 

Measurements of multi-jet cross sections have also been carried out in the analysis. Dijet distributions of figure \ref{fig7:progression_dijets} are calculated with the average $p_\mathrm{T}$ of the hardest two jets. Merging of two additional jets is again sufficient, with additional samples contributing only little to the cross section. The merging setups with fixed $\Tilde{t}_\mathrm{MS}$
fail to describe the $p_{\mathrm{T}}$-slope of the data, overshooting it in the low-$p_\mathrm{T}$ region and undershooting in the high-$p_\mathrm{T}$ region.
The CKKW-L merging with dynamic merging scale, instead, provides a good description of the data 
throughout the considered $p_{\mathrm{T}}$ range,
though some statistical fluctuations are still present. 

Trijet distributions in figure \ref{fig8:progression_trijets} are obtained as the average $p_\mathrm{T}$ of the three hardest jets. The main takeaway of these comparisons is that $+2$-jet merging with three partons in the final state is still not enough to describe trijets, and there is an improvement in the cross sections when going from $+2$-jet to $+3$-jet merging. In general, we find that accurately describing $N$-jet cross sections requires at least $(N+1)$-jet merging. The highest multiplicity sample in a given merging run is treated differently than the other samples, as described in section~\ref{sec2.2:merging}: the events of $+3$-jet sample are assigned more vetoes when doing $+4$-jet merging (or higher). This together with the smaller size of the $+4$-parton sample creates some statistical uncertainty. The resulting distributions may vary up or down compared to the previous sample, even in the CKKW-L algorithm which only has positively weighted events. The observation is clear in figures \ref{fig6:progression_incjets} and \ref{fig7:progression_dijets}. This kind of uncertainty is not accounted for in the upcoming figures, but could perhaps be reduced simply by increasing statistics of the last multiplicity sample, especially in the phase space in question. This, however, would be a significant computing effort.

\subsection{Merging scale variations}
\label{sec4.2:MergingScales}

\begin{figure}
    \centering
    \caption*{a)}
    \vspace{-2.5em}
    \begin{minipage}{.45\textwidth}
    \includegraphics[scale=0.55]{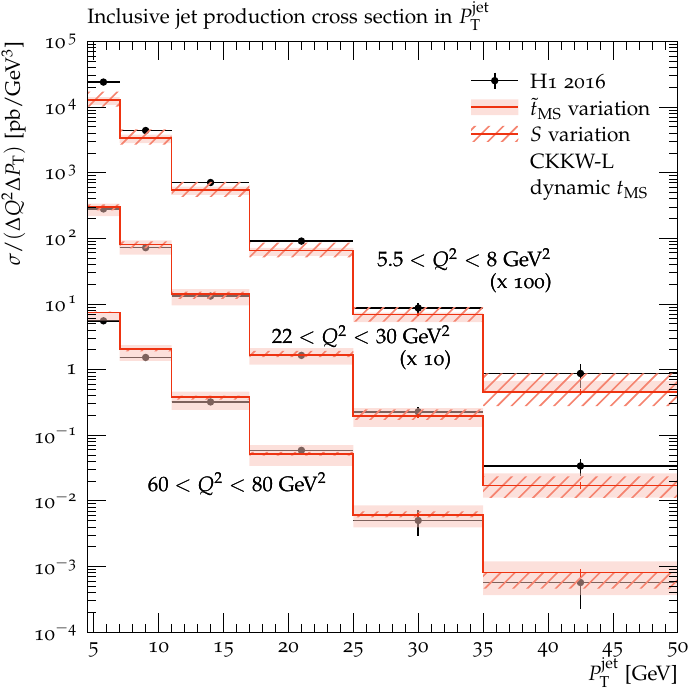}
    \end{minipage}
    \begin{minipage}{.45\textwidth}
    \vspace{1.2em}
    \hspace{-1.5em}
    \includegraphics[scale=0.6]{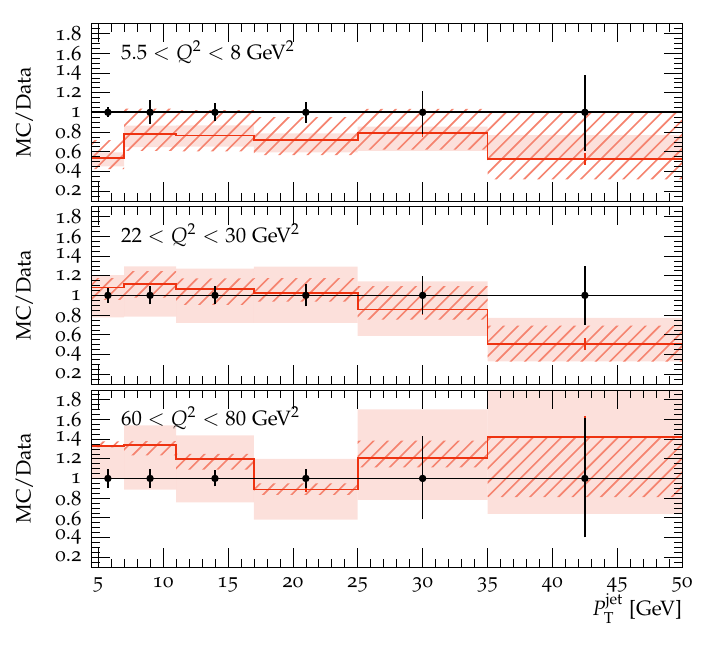}
    \end{minipage}

    \caption*{b)}
    \vspace{-2.5em}
    \begin{minipage}{.45\textwidth}
    \includegraphics[scale=0.55]{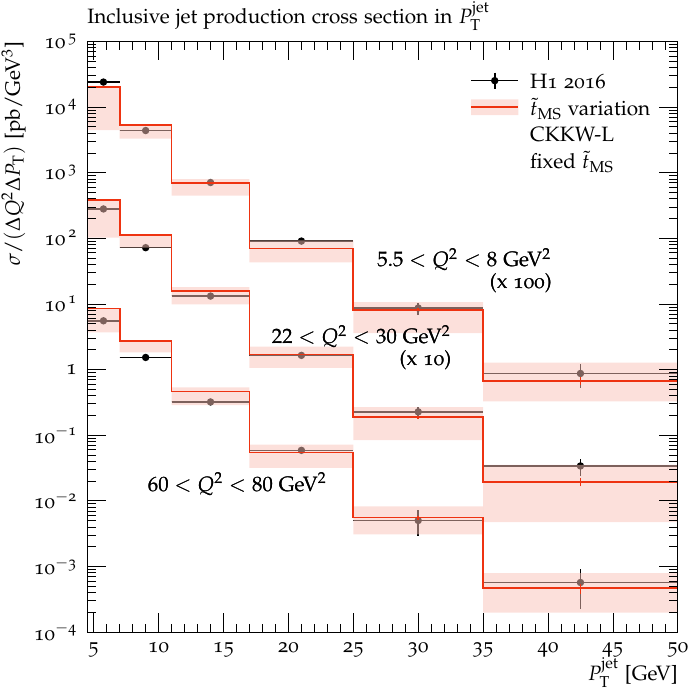}
    \end{minipage}
    \begin{minipage}{.45\textwidth}
    \vspace{1.2em}
    \hspace{-1.5em}
    \includegraphics[scale=0.6]{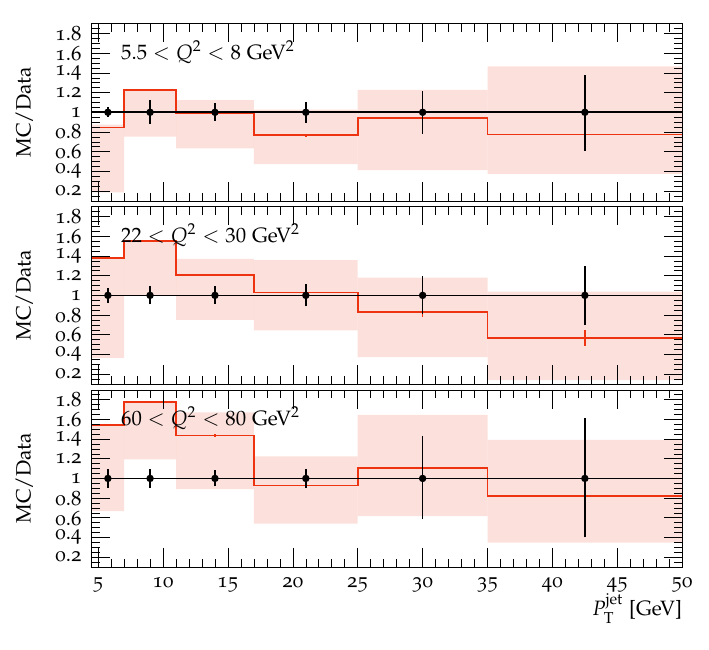}
    \end{minipage}

    \caption*{c)}
    \vspace{-2.5em}
    \begin{minipage}{.45\textwidth}
    \includegraphics[scale=0.55]{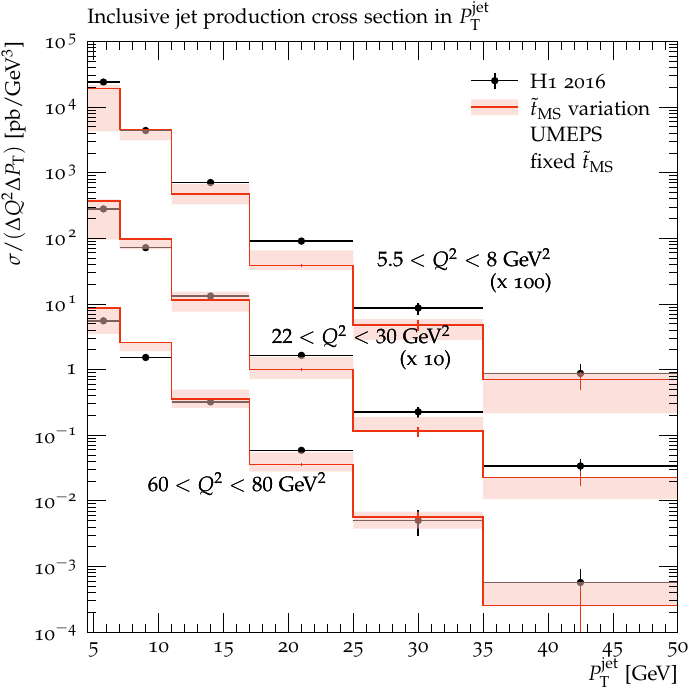}
    \end{minipage}
    \begin{minipage}{.45\textwidth}
    \vspace{1.2em}
    \hspace{-1.5em}
    \includegraphics[scale=0.6]{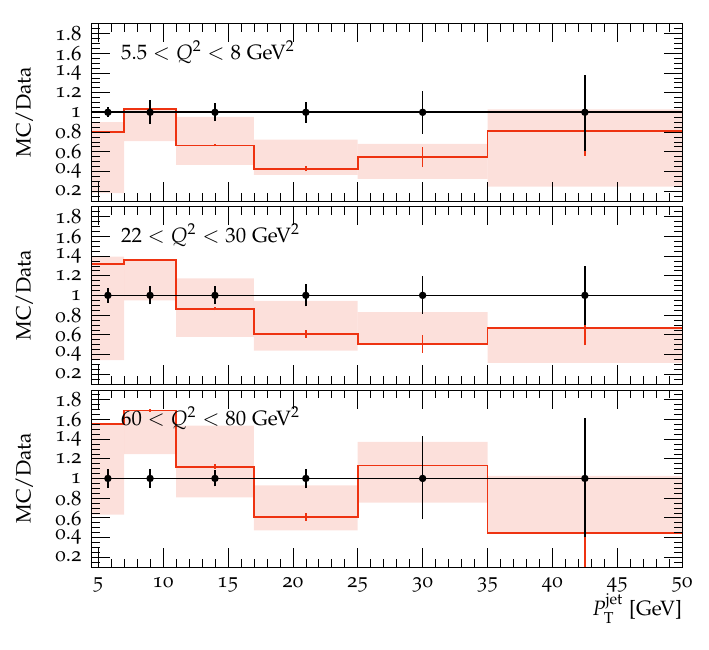}
    \end{minipage}

    \caption{Inclusive jet cross sections with 4 additional jets in a) dynamic CKKW-L merging with $\mu_\mathrm{F}^2 = Q^2$, b) fixed CKKW-L merging with $\mu_\mathrm{F}^2 = \frac{1}{2}\left(Q^2 + H_\mathrm{T}^2 \right)$ and c) fixed UMEPS merging with $\mu_\mathrm{F}^2 = \frac{1}{2}\left(Q^2 + H_\mathrm{T}^2 \right)$. Light-red error bands show the variation in $\Tilde{t}_{MS} = (5^{+4}_{-2}~\mathrm{GeV})^2$, and the diagonal line pattern shows the variation in parameter $S = 0.7 \pm 0.1$. Statistical uncertainties are shown as a vertical bar in each bin. Data and MC results have been scaled by a factor of $100$ for $5.5<Q^2<8$ GeV$^2$ and $10$ for $22<Q^2<30$ GeV$^2$. Data from H1 \cite{H1:2016goa}.}
    \label{fig9:IncJets}
\end{figure}

\begin{figure}
    \centering
    \caption*{a)}
    \vspace{-2.5em}
    \begin{minipage}{.45\textwidth}
    \includegraphics[scale=0.55]{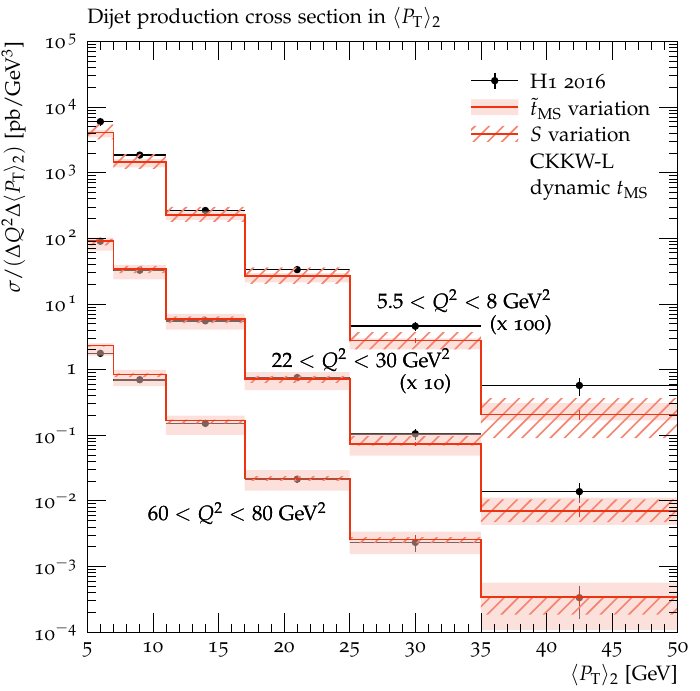}
    \end{minipage}
    \begin{minipage}{.45\textwidth}
    \vspace{1.2em}
    \hspace{-1.5em}
    \includegraphics[scale=0.6]{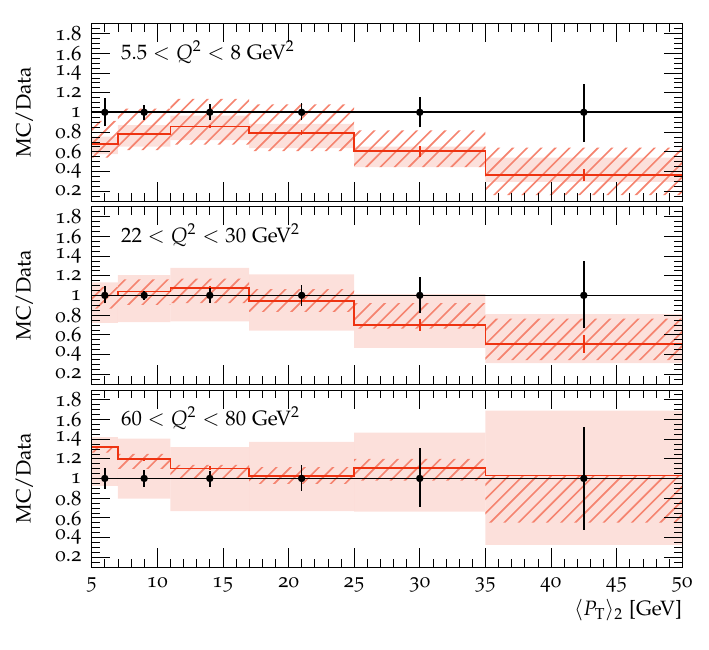}
    \end{minipage}

    \caption*{b)}
    \vspace{-2.5em}
    \begin{minipage}{.45\textwidth}
    \includegraphics[scale=0.55]{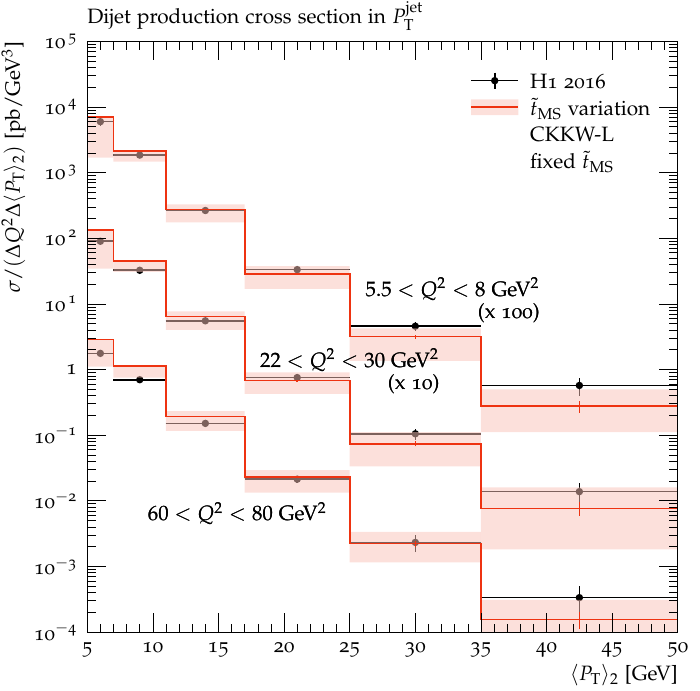}
    \end{minipage}
    \begin{minipage}{.45\textwidth}
    \vspace{1.2em}
    \hspace{-1.5em}
    \includegraphics[scale=0.6]{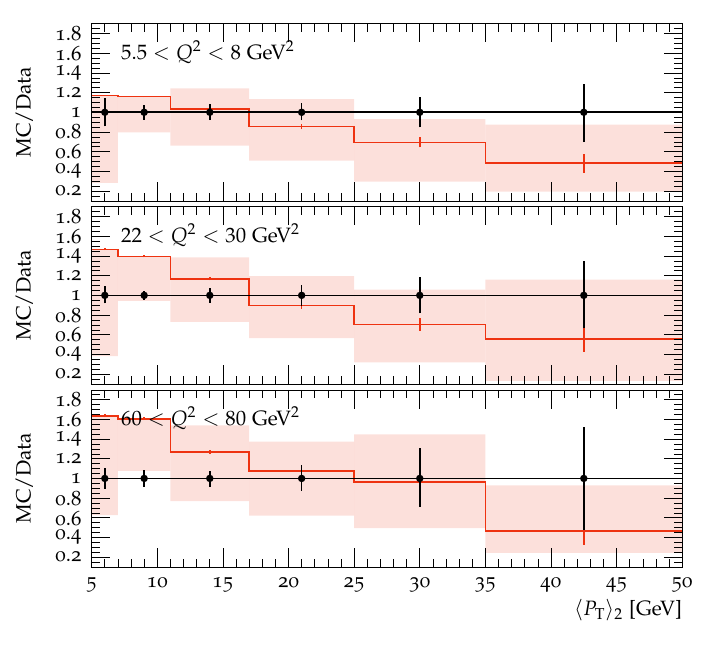}
    \end{minipage}

    \caption*{c)}
    \vspace{-2.5em}
    \begin{minipage}{.45\textwidth}
    \includegraphics[scale=0.55]{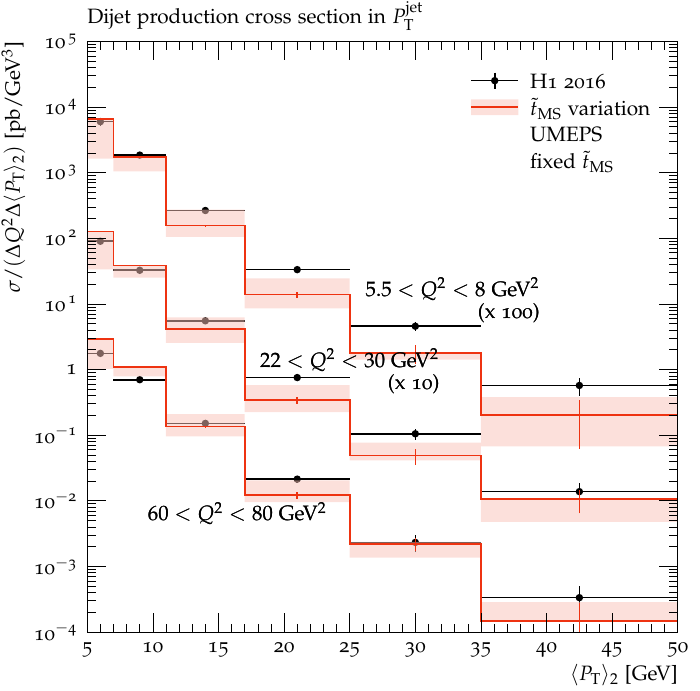}
    \end{minipage}
    \begin{minipage}{.45\textwidth}
    \vspace{1.2em}
    \hspace{-1.5em}
    \includegraphics[scale=0.6]{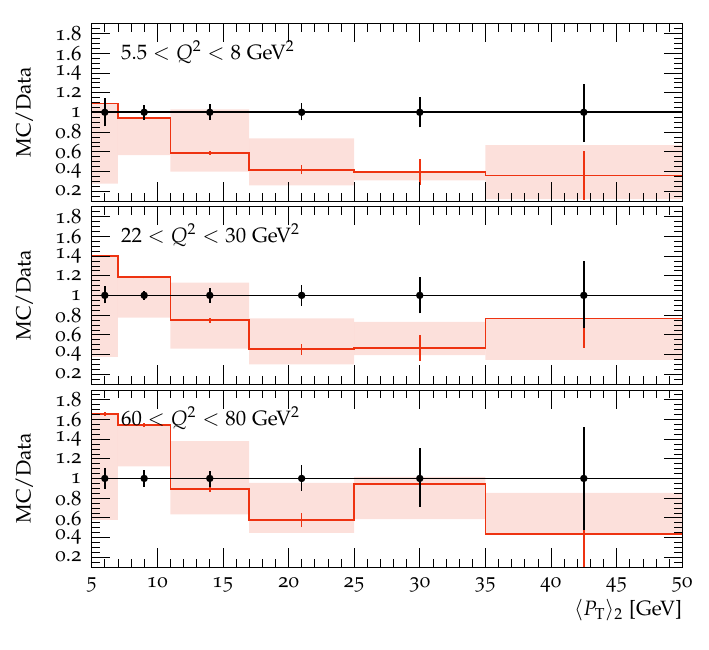}
    \end{minipage}

    \caption{Dijet cross sections of the three merging setups. Data from H1 \cite{H1:2016goa}. Details of the plot as in figure \ref{fig9:IncJets}. }
    \label{fig10:DiJets}
\end{figure}

\begin{figure}
    \centering
    \caption*{a)}
    \vspace{-2.5em}
    \begin{minipage}{.45\textwidth}
    \includegraphics[scale=0.55]{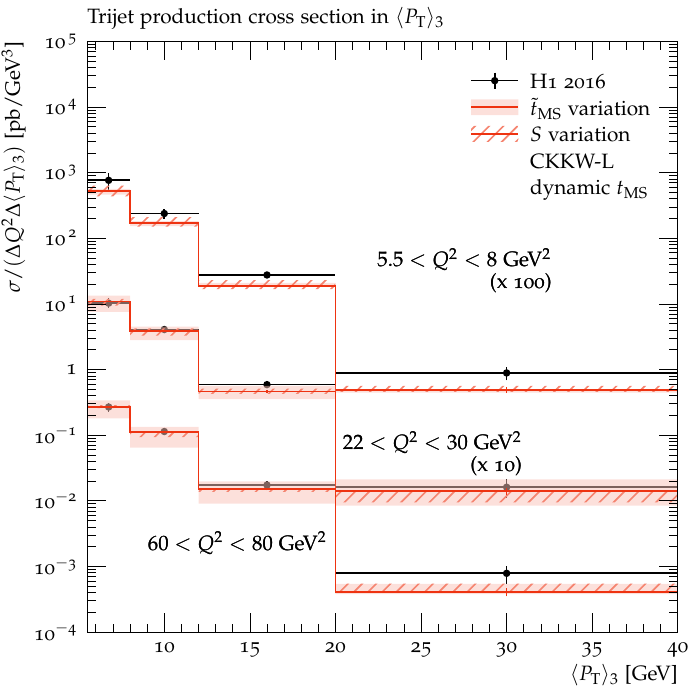}
    \end{minipage}
    \begin{minipage}{.45\textwidth}
    \vspace{1.2em}
    \hspace{-1.5em}
    \includegraphics[scale=0.6]{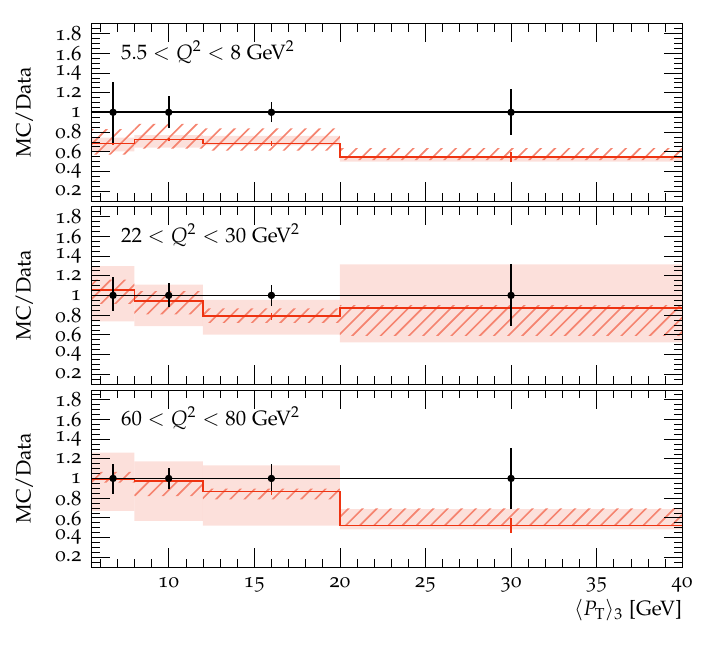}
    \end{minipage}

    \caption*{b)}
    \vspace{-2.5em}
    \begin{minipage}{.45\textwidth}
    \includegraphics[scale=0.55]{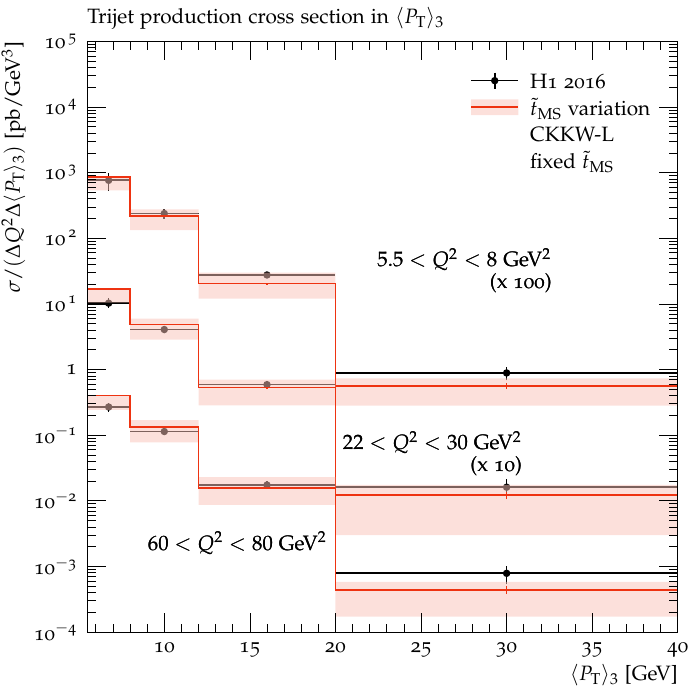}
    \end{minipage}
    \begin{minipage}{.45\textwidth}
    \vspace{1.2em}
    \hspace{-1.5em}
    \includegraphics[scale=0.6]{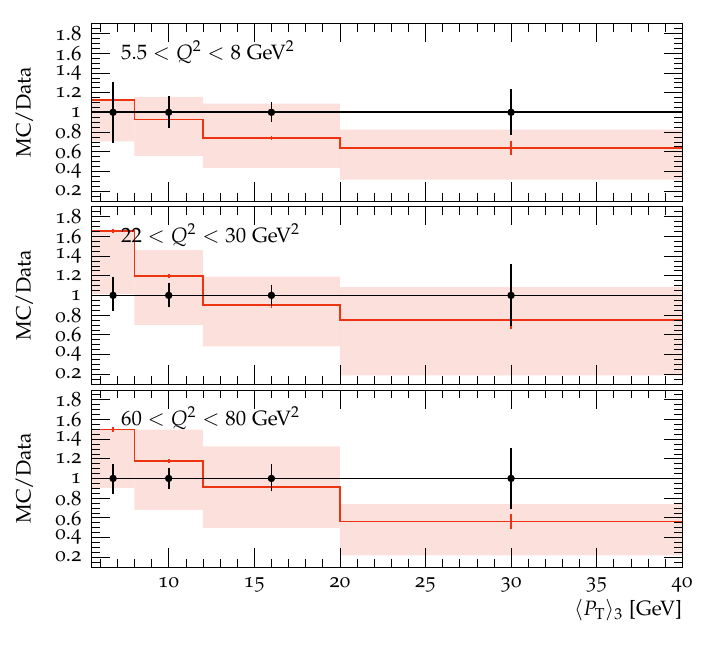}
    \end{minipage}

    \caption*{c)}
    \vspace{-2.5em}
    \begin{minipage}{.45\textwidth}
    \includegraphics[scale=0.55]{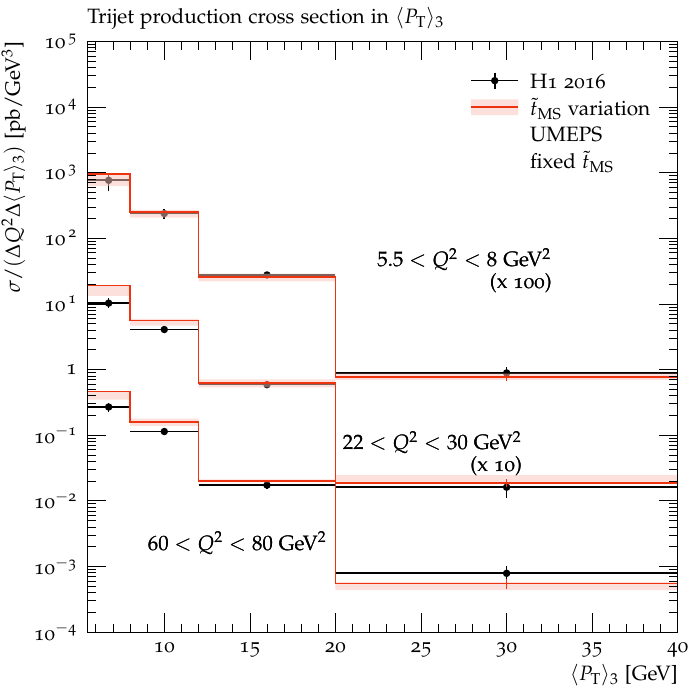}
    \end{minipage}
    \begin{minipage}{.45\textwidth}
    \vspace{1.2em}
    \hspace{-1.5em}
    \includegraphics[scale=0.6]{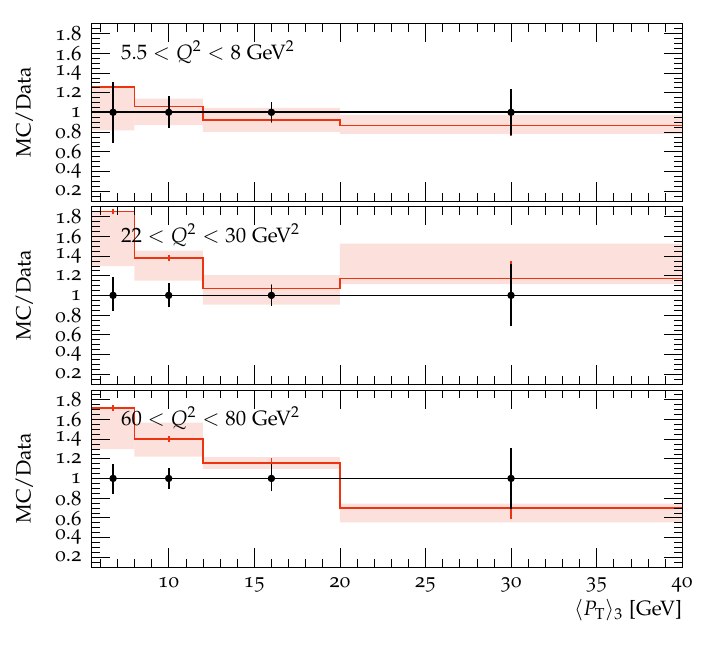}
    \end{minipage}

    \caption{Trijet cross sections of the three merging setups. Data from H1 \cite{H1:2016goa}. Details of the plot given in figure \ref{fig9:IncJets}. }
    \label{fig11:TriJets}
\end{figure}

For a sufficient description of jets at very low virtualities, a dynamic merging scale allows the matrix element contributions to access a larger fraction of phase space than in case of a fixed merging scale. The effect of varying the merging scale is presented in figures \ref{fig9:IncJets}-\ref{fig11:TriJets} for each of the three setups for three different $Q^2$ bins ranging from $5.5 ~\mathrm{GeV}^2$ to $80 ~\mathrm{GeV}^2$ for inclusive jets, dijets and trijets. The resulting cross sections are somewhat sensitive to parameter values related to the merging scale. In case of dynamic merging scale, panels a), we vary $S$ and $\Tilde{t}_\mathrm{MS}$ of eq. \eqref{eq2.26:DynamicTMS} within ranges of $S = 0.7 \pm 0.1$ and $\Tilde{t}_\mathrm{MS} = (5^{+4}_{-2}~\mathrm{GeV})^2$. Variation of the $S$-parameter has a 20\% effect in the inclusive jet cross section but is reduced to less than 10\% in higher virtuality events where the value of $t_\mathrm{MS}$ is close to $\Tilde{t}_\mathrm{MS}$, as evident from figure \ref{fig5:DynamicMergingScale}. Variation of the $\Tilde{t}_\mathrm{MS}$-parameter, in turn, has smaller impact at low-$Q^2$ bins but becomes dominant at larger values of $Q^2$. On average, the combined effect from these variations are around $40$\% for inclusive jet cross section and dijets and slightly smaller for trijets.

Panels b) of figures \ref{fig9:IncJets}--\ref{fig11:TriJets} depict the results for CKKW-L with fixed merging scale. Now the effect from $\Tilde{t}_\mathrm{MS}$ variation is increased but of similar order as the combined variation with the dynamical merging-scale option. The resulting cross sections are fairly similar as in case of the dynamic merging scale. At smaller values of $Q^2$ and $p_{\mathrm{T}}$ the fixed merging scale version gives about 20\% higher cross sections but the differences become smaller with increasing $Q^2$ and $p_{\mathrm{T}}$. To some extend this is expected since at higher $Q^2$ the $t_\mathrm{MS}$ approaches $\Tilde{t}_\mathrm{MS}$ by construction. The agreement with the data at low $p_{\mathrm{T}}$ seem to be slightly better with the dynamic merging scale.

Results with UMEPS merging using a fixed merging scale are compared to data in panels c) of figures \ref{fig9:IncJets}-\ref{fig11:TriJets}. It is apparent from the plots that the merging-scale variation is rather moderate in UMEPS merging compared to the two CKKW-L merging setups. This is a direct consequence of the more consistent scale treatment in UMEPS, as reflected in its add-and-subtract approach to merging. While being more stable against merging-scale variations, the agreement with the data is not as good as in case of CKKW-L as the cross sections seem to miss part of the high-$p_{\mathrm{T}}$ jets especially in the lower virtuality bins. Only in case of trijets, figure \ref{fig11:TriJets}, the agreement with the data is similar as with CKKW-L.

\subsection{Factorization and renormalization scale variations}
\label{sec4.3:MG5andSherpa}

\begin{figure}
    \centering
    \begin{minipage}{.45\textwidth}
    \includegraphics[scale=0.55]{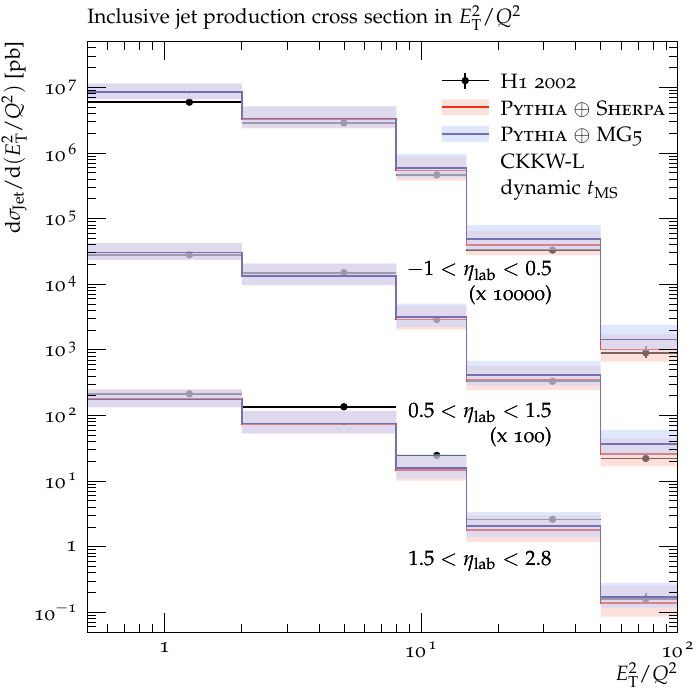}
    \end{minipage}
    \begin{minipage}{.45\textwidth}
    \vspace{1.2em}
    \hspace{-1.5em}
    \includegraphics[scale=0.6]{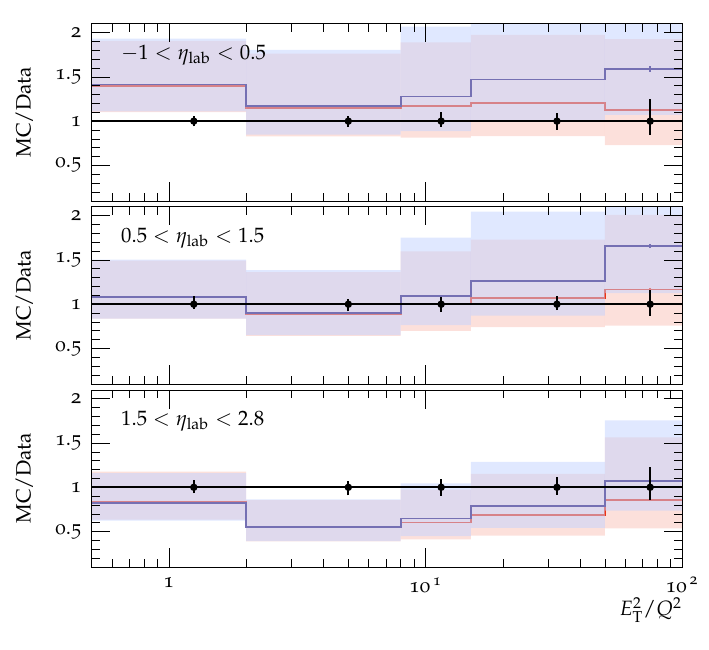}
    \end{minipage}
    \caption{Comparison of $+2$-jet CKKW-L dynamic merging on parton level events from \sherpa and \madgraph. Jet cross sections in terms of jet transverse energy by photon virtuality in different pseudorapidity regions. Uncertainty bands from factorization and renormalization scale variations. Data from H1 \cite{H1:2002qhb}. }
    \label{fig12:CKKWL_eta}
\end{figure}

Figure \ref{fig12:CKKWL_eta} shows the comparison of $+2$-jet merging with parton-level events from \madgraph and \sherpa. We expect $+2$-jet merging to be sufficient for inclusive-jet cross sections, as discussed in section~\ref{sec4.1:JetMultiplicities}. The  uncertainty bands are the standard $7$-point variations of factorization and renormalization scales, obtained by multiplying or dividing the scales by 2. Describing measured jet cross sections in terms of pseudorapidity $\eta$ proved challenging in our multi-jet merging implementation. We achieve a good description of central rapidity region in both cases, but the distributions of figure \ref{fig12:CKKWL_eta} are systematically too high for the backwards region and too low for the forward region, but still within uncertainties. These cross sections are differential in jet transverse energy squared over virtuality, which is a suitable observable for jet studies, especially in case of multi-jet merging. The observable captures the low-$Q^2$, high-$E_\mathrm{T}^2$ dynamics well, the region which most benefits from multi-jet merging.
The comparisons show a good agreement between approaches using hard events from \madgraph and \sherpa, also for the uncertainties from scale variations, but their results diverge in the high-end of $E_\mathrm{T}^2/Q^2$.

\newpage
\section{Conclusion}
\label{sec5:conclusion}

In this work, we have presented an implementation of multi-jet merging in DIS with \pythia, using the \vincia antenna-shower framework. We have employed the CKKW-L and UMEPS algorithms with parton-level event input from \sherpa and cross-checked the results using hard processes from \madgraph. We have studied the effects of different factorization and renormalization-scale choices along with different merging-scale prescriptions. We found that cross sections for $N$-jet production started to level off when we included events up to $N+1$ partons in the merging algorithms indicating that it is important to account also for sub-leading jet kinematics when studying jet production in DIS. Comparing the results to experimental data from the H1 collaboration, we find good agreement of differential multi-jet cross sections across a wide range of photon virtualities.

In case of CKKW-L merging we studied two different merging scales, a fixed one and a $Q^2$-dependent option referred to as dynamic merging scale. In case of the former the renormalization and factorization scales accounted also for the transverse momentum of the produced partons to partly capture the idea of the dynamic merging scale where the part of the phase space with high-$p_{\mathrm{T}}$ partons and low $Q^2$ would be mainly covered by the hard partons instead of parton shower. Indeed, we found that the resulting cross sections were quite similar with these approaches. However, we found that the fixed merging scale lead to an increased sensitivity to the value of the merging scale compared to the dynamic setting and that the resulting $p_{\mathrm{T}}$-slope was slightly steeper. In case of UMEPS we could only use the fixed merging scale option. This algorithm lead to a more stable cross sections against the merging scale variations. However, the agreement with the H1 data was somewhat worse than with CKKW-L for inclusive and dijet production at high $p_{\mathrm{T}}$.

In case of inclusive and dijet production we noticed that it was necessary to include hard events with three partons in the final states to get converging cross sections and a good agreement with the H1 data. In case of inclusive jet cross section this might be somewhat surprising but it is good to keep in mind that in the Breit frame the Born-level events will not contain any outgoing partons at high $p_{\mathrm{T}}$. In case of tri-jet production, the higher-multiplicity parton-level events still improved the description of the data, especially with merging prescriptions using a fixed merging-scale. These observations indicate that for modelling jet production in DIS at moderate $Q^2$, it is important to account also for the kinematics of the sub-leading jets as was pointed out in ref.~\cite{Carli_2010}.

To have a handle on the theoretical uncertainties related to missing higher-order corrections, we considered variations of renormalization and factorization scales in case of CKKW-L with dynamic derivation of the merging scale. We found that such variations typically lead to 30-50\% modifications of the final cross sections. As a cross check, we have also considered the use of events generated with \madgraph, and found a very good agreement with events from \sherpa after merging, though with a slight mismatch at the highest $E_{\mathrm{T}}^2/Q^2$ bins. Comparing the outcomes of the three merging setups over the full range of observables considered here, we conclude that the recommended setup is the CKKW-L algorithm with a dynamic merging scale and the renormalization and factorization scales set to the photon virtuality. This setup proved most consistent with data of multi-jet cross sections and being least sensitive to merging-scale variations in case of jet rates.

The DIS-specific multi-jet merging implementation described in this work will be made publicly available as part of an upcoming release of \pythia~8.3, with the possibility to use event files in the LHEF and HDF5 formats. Our work marks an important step to enable well-founded multi-jet event simulations for the EIC. Some prospects for further improvements in modelling DIS with \pythia are the extension of the multi-jet merging framework to next-to-leading order, consistent inclusion of photoproduction events and an implementation of merging capabilities with other parton shower algorithms in \pythia. A dynamic merging-scale prescription in the UMEPS implementation presented here may also be studied further. As an intermediate step, a consistent internal NLO matching scheme may be developed, based on \pythia's original matrix-element correction framework.

\acknowledgments
We would like to express our gratitude to Frank Krauss, Leif Lönnblad and Olivier Mattelaer for their insightful discussions and valuable answers regarding this study. We would like to thank Stefan Höche for comments on the DIS merging implementation in \sherpa and help with the scale setting in the HPC-enabled version of \sherpa.
We are grateful to Stephen Mrenna for comments on the manuscript.
We acknowledge grants of computer capacity from the Finnish Grid and Cloud Infrastructure (persistent identifier urn:nbn:fi:research-infras-2016072533).
Part of the computations were carried out on the PLEIADES cluster at the University of Wuppertal, supported by the Deutsche Forschungsgemeinschaft (DFG, grant No. INST 218/78-1 FUGG) and the Bundesministerium für Bildung und Forschung (BMBF).

\paragraph{Funding information}
This research was funded through the Research Council of Finland,
projects 330448, 331545, 336419 and 361179, and was a part of the
Center of Excellence in Quark Matter of the Research Council of
Finland, project 346326. This research is part of the European Research
Council project ERC-2018-ADG-835105 YoctoLHC.








\bibliographystyle{JHEP}
\bibliography{DISJetMerging}

\end{document}